\documentclass[useAMS,usenatbib]{mn2e}
\usepackage[letterpaper,margin=0.65in]{geometry}
\usepackage[ansinew]{inputenc}
\usepackage{amsfonts}
\usepackage{amsmath}
\usepackage{enumerate}
\usepackage{enumitem}
\usepackage{graphicx}
\usepackage{epsfig}
\usepackage{color}
\usepackage{amssymb}

\newcommand{\eg}{{e.g., }}
\newcommand{\ie}{{i.e., }}
\newcommand{\pz}{photo-$z$\ }
\newcommand{\pzs}{photo-$z$s }
\newcommand{\tpz}{\texttt{TPZ} }
\newcommand{\somz}{\texttt{SOM$z$} }
\newcommand{\bpz}{\texttt{BPZ} }

\newcommand{\PZ}{Photo-$z$\ }

\newcommand{\tpzns}{\texttt{TPZ}}
\newcommand{\somzns}{\texttt{SOM$z$}}
\newcommand{\bpzns}{\texttt{BPZ}}
\newcommand{\pzns}{photo-$z$} 

\renewcommand{\vec}[1]{\mathbf{#1}}

\title[On how to efficiently combine photo-$z$ PDFs]
{Exhausting the Information: Novel Bayesian Combination of Photometric Redshift PDFs}
\author[M. Carrasco Kind and R. J. Brunner] 
{Matias Carrasco Kind\thanks{E-mail: mcarras2@illinois.edu} and Robert J. Brunner\\
Department of  Astronomy, University of Illinois, Urbana, IL 61820 USA}

\begin{document}
\date{\today}

\pagerange{\pageref{firstpage}--\pageref{lastpage}} \pubyear{2014}

\maketitle

\label{firstpage}
\begin{abstract}

The estimation and utilization of photometric redshift probability density functions (photo-$z$ PDFs) has become increasingly important over the last few years and currently there exist a wide variety of algorithms to compute photo-$z$'s, each with their own  strengths and weaknesses. In this paper, we present a novel and efficient Bayesian framework that combines the results from different photo-$z$ techniques into a more powerful and robust  estimate by maximizing the information from the photometric data. To demonstrate this we use a supervised machine learning technique based on random forest, an unsupervised method based on self-organizing maps, and a standard template fitting method but can be easily extend to other existing techniques. We use data from the DEEP2 and the SDSS surveys to explore different methods for combining the predictions from these techniques. By using different performance metrics, we demonstrate that we can improve the accuracy of our final photo-$z$ estimate over the best input 
technique, that the fraction of outliers is reduced, and that the identification of outliers is significantly improved when we apply a Na\"{\i}ve Bayes Classifier to this combined  information. Our more robust and accurate photo-$z$ PDFs will allow even more precise cosmological constraints to be made by using current and future photometric surveys. These improvements are crucial as we move to analyze photometric data that push to or even past the limits of the available training data, which will be the case with the Large Synoptic Survey Telescope.

\end{abstract}


\section{Introduction}

Spectroscopic galaxy surveys have played an important role in understanding the origin, composition, and evolution of our Universe. Surveys like the Sloan Digital Sky Survey (SDSS;~\citealt{York2000}), WiggleZ~\citep{Drinkwater2010}, and BOSS~\citep{Dawson2013} have imposed important constraints on the allowed parameter values of the standard cosmological model ~\citep[\eg][]{Percival2010,Blake2011,Sanchez2013}. However, spectroscopic measurements are considerable more expensive to obtain than photometric data, they are more likely to suffer from selection effects, and they provide much smaller galaxy samples per unit telescope time. As a consequence, current ongoing and future galaxy surveys like the Dark Energy Survey (DES\footnote{http://www.darkenergysurvey.org/}) and the Large Synoptic Survey Telescope (LSST\footnote{http://www.lsst.org/lsst/}) are pure photometric surveys. These surveys will  enable cosmological measurements on galaxy samples that are currently at least a hundred times larger than 
comparable spectroscopic samples, that have relatively simple and uniform selection functions, that extend to fainter flux limits and larger angular scales, thereby probing much larger cosmic volumes and will photometrically detect galaxies that are too faint to be spectroscopically observed. 

With the growth of these large  photometric surveys, the estimation of galaxy redshifts by using multi band photometry has grown significantly over the last two decades. As a result, a variety of different algorithms for estimating \pz's based on statistical techniques  have been developed ~\citep[see, \eg][for a review of current \pz techniques]{Hildebrandt2010,Abdalla2011,Sanchez2014}. Over the last several years, particular attention has been focused on techniques that compute a full probability density function (PDF) for each galaxy in the sample. A \pz PDF contains more information than a single \pz estimate, and the use of \pz PDFs has been shown to improve the accuracy of cosmological measurements ~\citep[\eg][]{Mandelbaum2008,Myers2009,Jee2013}.

\PZ techniques  can be broadly divided into two categories: spectral energy distribution (SED) fitting, and training based algorithms. Template fitting approaches \citep[see \eg][]{Benitez2000,Bolzonella2000,Feldmann2006,Ilbert2006,Assef2010} estimate \pzns s by finding the best match between the observed set of magnitudes or colors, and the synthetic magnitudes or colors taken from the suite of templates that are sampled across the expected redshift range of the photometric observations. This method is often preferred over empirical techniques as they can be applied without obtaining a high-quality spectroscopic training sample. However, these techniques do require a representative sample of template galaxy spectra, and they are not exempt from uncertainties due to measurement errors on the survey filter transmission curves, mismatches when fitting the observed magnitudes or colors to template SEDs, and color--redshift degeneracies. The use of training data that include known redshifts can also improve 
these 
predictions~\citep[\eg][]{Ilbert2006, Newman2013b}.  On the other hand, machine learning methods have been shown to have similar or even better performance~\citep[\eg][]{Collister2004, CarrascoKind2013a} when the spectroscopic training sample is populated by representative galaxies from the photometric sample.

Machine learning methods have the advantage that it is easier to include extra information, such as galaxy profiles, concentrations, or different modeled magnitudes within the algorithm. However, they are only reliable within the limits of the training data, and one must exercise sufficient caution when extrapolating these algorithms. These techniques can be sub-categorized into supervised and unsupervised machine learning approaches. For supervised techniques~\citep[\eg][]{Connolly1995,Brunner1997, Collister2004,Wadadekar2005,Ball2008,Lima2008,Freeman2009,Gerdes2010, CarrascoKind2013a}, the input attributes (e.g., magnitudes or colors) are provided along with the desired output (e.g., redshift). This training information is  
directly used by the algorithm during the learning process. In this case, the redshift information from the training set \textit{supervises} the learning process and decisions are made by using this information. On the other hand,  unsupervised machine learning \pz techniques \citep[\eg][]{Geach2012,Way2012, CarrascoKind2014a} are less common as they do not use the desired output value (e.g., redshifts from the spectroscopic sample) during the training process. Only the input attributes are processed during the training, leaving aside the redshift information until the evaluation phase.

Given the importance of these \pz PDFs, there is a present demand to compute them as efficiently and accurately as possible. Additional requirements include the need to understand the impact of systematics from the spectroscopic sample on the estimation of these  PDFs~\citep[\eg][]{Oyaizu2008,Cunha2012a,Cunha2012b}, and to maximally reduce  the fraction of catastrophic outliers~\citep[\eg][]{Gorecki2014}. Considerable effort has, therefore, been put into both the development of different techniques and the exploration of new approaches in order to maximize the efficacy of \pz PDF estimation. Yet, the combination of multiple, independent \pz PDF techniques has remained under explored~\citep[\eg][]{CarrascoKind2013b, Dahlen2013}. 

In this paper we extend our previous exploratory work in combining machine learning techniques with template fitting methods~\citep{CarrascoKind2013b}  to explicitly address this issue by presenting a novel Bayesian framework to combine and fully exploit different \pz PDF techniques. In particular, we show that the combination of a standard template fitting technique with both a supervised and an unsupervised machine learning method can improve the overall accuracy over any individual method. We also demonstrate how this combined approach can both reduce the number of outliers 
and improve the identification of catastrophic outliers when compared to the individual techniques. Finally, we show that this methodology can be easily extended to include additional, independent techniques and that we can maximize the complex information contained within a photometric galaxy sample.

This  paper is organized as follows.  In Section 2 we present the algorithms used in this work to generate the individual  \pz PDF estimates and we provide a brief description on their individual functionality. We describe, in Section 3, the different Bayesian approaches by which different \pz techniques are combined. Section 4 introduces the data sets employed to test this Bayesian approach taken from the SDSS and DEEP2 surveys. In Section 5 we present the main results of our combination approach and compare these results to those from the individual \pz PDF methods. In Section 6 we discuss the application of a Na\"{\i}ve Bayes combination technique for outlier detection. In Section 7 we conclude with a summary of our main points and a more general discussion of this new approach.

\section{Photo-z methods}\label{pz_methods}

To develop and test our combination framework, we consider three, distinct  \pz PDF estimation techniques; we briefly discuss each one of them in this section. We make the reasonable assumption that these three techniques are independent in their nature where two of these methods implement machine learning algorithms. The first method is a supervised machine learning technique we have published called \tpz ~\citep[Trees for Photo-Z,][hereafter CB13]{CarrascoKind2013a}, which uses prediction trees and a random forest to produce probability density functions. The second method is an unsupervised technique we have published called \somz ~\citep[][hereafter CB14]{CarrascoKind2014a}, which uses self organizing maps (SOM) and a random atlas to produce a probability density function. We have recently incorporated these two implementations into a new, publicly available and growing \pz PDF prediction framework called \texttt{MLZ}\footnote{\label{fnote} http://lcdm.astro.illinois.edu/code/mlz.html} (Machine Learning 
for photo-Z).

The third method is a Bayesian template fitting technique based on \bpz~\citep[Bayesian Photometric Redshifts;][]{Benitez2000}, which fits spectral energy density templates from a preselected  library to an observed set of measured flux values. Taken together, these three methods span the three standard published approaches in computing \pzns s in the literature. Any new method would, very likely, be functionally similar to one of these three methods; therefore, any of these three methods could in principle be replaced by a similar method to avoid redundancy. This can be most easily demonstrated for template fitting methods, where an additional set of \pz estimations can be utilized by adopting a different template library~\cite[e.g.,][]{Dahlen2013}. In this particular case, the underlying code is essentially unchanged, but the \pz results will change as different spectral libraries are adopted.

\begin{figure*}
\includegraphics[width=0.39\textwidth]{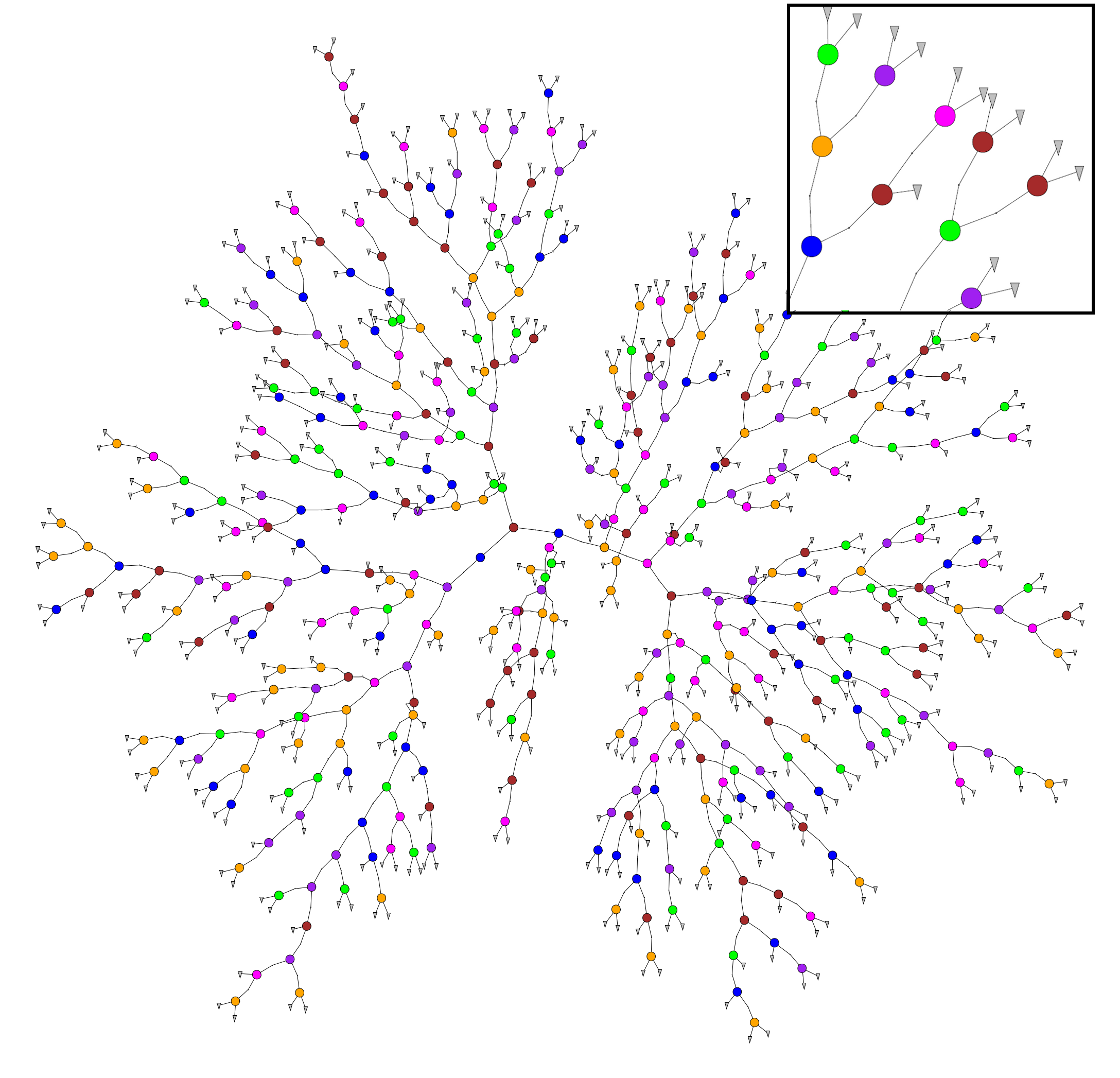}
\includegraphics[width=0.49\textwidth]{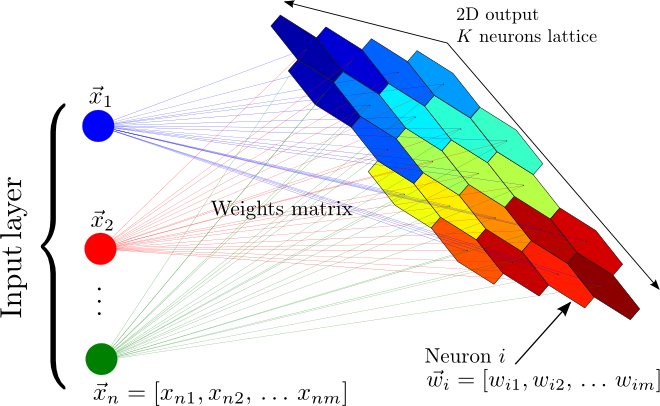}
\caption{\textit{Left}: A simplified example of a  binary prediction tree plotted radially, taken from CB13. The initial node is close to the center of the figure; each node is subdivided and the splitting process terminates when a pre-defined stopping criterion is reached. Individual colors represent a unique variable (\eg a magnitude like $g$ or $r$, or a color like $g - r$) used to split an individual node. Each leaf node provides a specific prediction based on the information contained within that terminal node (gray triangles in the figure). The subpanel highlights a specific branch of the tree at higher resolution for additional clarity. \textit{Right}: A schematic representation of a self organized map, taken from CB14. The training set of $n$ galaxies is mapped onto a two-dimensional lattice of $K$ neurons that are represented by vectors containing the weights for each input attribute. Note that the galaxies and the weight vectors are of the same dimension $m$, and that one neuron can represent more 
than 
one training galaxy. The colors used in the map encode the target property from the galaxies grouped within that cell.}
\label{fig:mlz}
\end{figure*}

\subsection{TPZ}
 \tpz (CB13) is a parallel, supervised algorithm that uses prediction trees and random forest techniques ~\citep{Breiman1984, Breiman2001} to produce \pz PDFs and ancillary information for a sample of galaxies. Among the different non-linear methods that are used to compute photometric redshifts, prediction trees and random forests are one of the simplest yet most accurate techniques. Furthermore, they have been shown to be one of the most accurate algorithms for low as well as high multi-dimensional data~\citep{Caruana2008}.

Prediction trees are built by asking a sequence of questions that recursively split the data into two branches until a terminal leaf is created that meets a pre-defined stopping criterion (\eg a minimum leaf size or a maximum rms within that leaf). The small region bounding the data in the terminal leaf node represents a specific subsample of the entire data that all share similar characteristics. A comprehensive predictive model is applied to the data within each leaf that enables predictions to be rapidly computed in situations where many variables might exist that possibly interact in a nonlinear manner, which is often the case with \pz estimation. A visualization of an example tree generated by \tpz is shown in the left panel of Figure~\ref{fig:mlz}. In this figure, the plotting colors represent the magnitudes (or source colors) in which the data are recursively divided. In practice, however, the prediction trees are generally both denser and deeper than the sample tree shown in the Figure.

To compute \pz PDFs in this study, we have used regression trees, which are a specific type of prediction trees. Regression trees are  built by first starting with a single node that encompasses the entire data, and subsequently  splitting the data within a node recursively into two branches along the dimension that provides the most information about the desired output. The procedure used to select the optimal split dimension is based on the minimization of the sum of the squared errors, which for a specific ${\rm node}$ is given by
 \begin{equation}\label{S_RT1}
 S({\rm node}) = \sum\limits_{m \in values(M)} \sum\limits_{i \in m} (z_i - \hat{z}_m)^2
 \end{equation}
where $m$ are the possible values (bins) of the dimension $M$, $z_i$ are the values of the target variable on each branch, and $\hat{z}_m$ is the specific prediction model used. In the case of the \textit{arithmetic mean}, for example, we would have that $\hat{z}_m = \frac{1}{n_m}\sum_{i \in m} z_i$, where $n_m$ are the members on branch $m$. This allows us to rewrite Equation \ref{S_RT1} as
\begin{equation}\label{S_RT2}
 S({\rm node}) = \sum\limits_{m \in values(M)} n_m V_m
\end{equation}
where $V_m$ is the variance of the estimator $\hat{z}_m$. 

At each node in our tree, we scan all dimensions to identify the split point that minimizes the function $S({\rm node})$. We choose the dimension that minimizes $S({\rm node})$ as the splitting direction, and this process is recursively repeated until either a predefined threshold in $S({\rm node})$ is reached or any new child nodes would contain less than the predefined minimum leaf size. When constructed, each terminal leaf within the prediction tree \textit{contains}  spectroscopic data with different redshift values; the final prediction value for a given leaf node is determined from a regression model that covers these spectroscopic data. The simplest model is to simply return the mean value of the set of spectroscopic training redshifts contained within the leaf node, which provides a single estimate of a continuous variable. Alternatively, all of the spectroscopic training redshifts can be retained and subsequently combined with data from the matching leaf nodes in other prediction trees to form an 
aggregate, final prediction.

We create bootstrap samples from the input training data by sampling repeatedly from the magnitude using the magnitude errors. We use these bootstrap samples to construct multiple, uncorrelated prediction trees whose individual predictions are aggregated to construct a \pz PDF for each individual galaxy by using a technique called a random forest.  We also use a cross validation technique called Out-of-Bag~\citep[][CB13]{Breiman1984} within \tpz to provide extra information about the galaxy sample. This information includes an unbiased estimation of the errors and a ranking of the relative importance of the individual input attributes used for the prediction. This extra information can prove extremely valuable when calibrating the algorithm, when deciding what attributes to incorporate in the construction of the forest, and when combining this approach with other techniques.

\tpz  has been tested extensively on different datasets, including the SDSS, DEEP2, and DES. In all tests,  \tpz has performed comparable to if not better than other machine learning approaches. When high quality training data are available, \tpz has been shown to actually outperform other comparable techniques, both training and template based. \cite{CarrascoKind2013a}  provides a more detailed discussion of the \tpz algorithm and its application to different datasets. 

\subsection{SOM$z$}

A Self Organized Map (SOM):~\citep{Kohonen1990,Kohonen2001} is an unsupervised, artificial neural network algorithm that is capable of projecting high-dimensional input data  onto a low-dimensional map through a process of competitive learning. In our case, the high dimensional input data can be galaxy magnitudes, colors, or some other photometric attributes, and  two dimensions are generally sufficient for the output map. A SOM differs from other neural network based-algorithms in that a SOM is unsupervised (the redshift information is not used during training), there are no hidden layers and therefore no extra parameters, and it produces a direct mapping between the training set and the output network. In fact, a SOM can be viewed as a non-linear generalization of a principal component analysis (PCA). 

The key characteristic of the self organization is that it retains the \textit{topology} of the input training set, revealing correlations between inputs that are not obvious. The method is unsupervised since the user is not required to specify the desired output during the creation of the low-dimensional map, as the \textit{mapping} of the components from the input vectors is a natural outcome of the competitive learning process. Another important characteristic of a SOM when applied to \pz estimation is the creation of a structured ordering of the spectroscopic training data, since similar galaxies in the training sample are mapped to neighboring neural nodes in the trained feature map (CB14).

We demonstrate the construction of a self-organizing map in the right-hand panel of Figure~\ref{fig:mlz}. During this phase, each node on the two-dimensional map is represented by weight vectors of the same dimension as the number of attributes used to create the map itself. In an iterative process, each galaxy in the input sample is individually used to correct these weight vectors. This correction is determined so that the specific neuron (or node), which at a given moment best represents the input galaxy, is modified along with the weight vectors of that node's neighboring neurons. As a result, this \textit{sector} within the map becomes a better representation of the current input galaxy. 

This process is repeated for every galaxy in the train sample, and this entire process is repeated for several iterations. Eventually the SOM converges to its final form where the training data is separated into \textit{groups} of similar features, which is illustrated in Figure~\ref{fig:mlz} by the different cell colors within the output map. The result of this direct mapping procedure is an approximation of the galaxy training probability density function, and the map itself can be considered a simplified representation of the full attribute space of the input galaxy sample. 

Building on our experience in creating \tpzns, we have developed a similar approach, named \somz (CB14), where prediction trees are replaced by SOMs to create what we called a \textit{random atlas}. The random atlas is constructed from multiple maps that are each constructed from different bootstrap samples selected from the input training data by perturbing the input attributes using their measured error, where each one of these maps are built using a random subsample of the attribute space. The multiple, uncorrelated maps are aggregated to generate a \pz PDF, in a similar manner as described earlier for the random forest. 

As described previously, our SOM implementation not only updates the best-matching node but also the topologically closest nodes to it. This functionality ensures that the entire region surrounding the best-matching node is identified as being similar to the current input galaxy. As a result, similar nodes within the map are co-located, which naturally mimics how the input galaxies that have similar properties tend to be co-located in the higher dimensional input parameter space. We apply this procedure iteratively to all input galaxies, which are processed randomly during each iteration to avoid any biases that might arise if galaxies are processed in a specific order.

When  running \somzns, there are few different parameters that must be determined, including the map resolution (\ie the number of pixels in the map), the number of iterations required to build the map, and, most importantly, the underlying two-dimensional topology used for the maps. In this paper we follow the guidelines we presented in CB14 for these parameters, and use a spherical topology for the map, which are constructed by using \texttt{HEALPIX}~\citep{Gorski2005}, where  each pixel in our maps has the same area. This topology was shown to be more accurate in many cases when compared to other topologies like a rectangular or hexagonal grid. In addition, a spherical topology  has natural periodic boundary conditions which avoids possible edge effects.

In analogy with \tpzns, we use cross validation, or OOB data, to estimate unbiased errors and to determine the relative importance of the different input attributes for this technique. These are both key pieces of information that will be used during the combination process, as we need to ensure that the same process is uniformly applied to each \pz estimation technique. By doing this, we will enable a robust analysis of the final results from the combination of the different techniques. \cite{CarrascoKind2014a} (CB14)  provides a complete description of the \somz implementation, the performance of this technique when applied to real data, and an exploration of specific parameter configurations.

\subsection{Template fitting approach}\label{template}

Using spectral templates to estimate galaxy \pzs from broadband photometry has a long history~\citep{Baum1962}; and this approach is, not surprisingly, one of the most utilized techniques. A primary advantage of this technique is the fact that a training sample is not required, thus this approach can be considered unsupervised. On the other hand, this technique has the disadvantage that a complete and representative library of spectral energy distributions (SEDs) are required. Thus any incompleteness in our knowledge of the template SEDs that fully span the input galaxy photometry will lead to inaccuracies or misestimates in the computation of a galaxy \pzns.

A number of different groups have published template fitting \pz estimation methods, all of which are roughly similar in nature. In this work, we have modified and parallelized one of the most popular, publicly available template fitting algorithms, \bpz~\citep{Benitez2000}. \bpz uses Bayesian inference to quantify the relative probability that each template matches the galaxy input photometry and determines a \pz PDF by computing the posterior probability that a given galaxy is at a particular redshift. We can write this probability as $P(z\mid\vec{x})$ for a specific template $t$, where  $\vec{x}$ represents a given set of magnitudes (or colors). If the identification of a specific template is not required, we can later marginalize over the entire set of templates $\vec{T}$. 

By using Bayes theorem, we have:
\begin{equation}
 P(z \mid \vec{x}) = \sum\limits_{t \in \vec{T}} P(z,t\mid\vec{x}) \propto \sum\limits_{t \in \vec{T}} \mathcal{L}(\vec{x} \mid z,t) P(z,t) .
\end{equation}
$\mathcal{L} (\vec{x} \mid z,t) $ is the likelihood that, for a given redshift $z$ and spectral template $t$, a specific galaxy has the set of magnitudes (or colors) $\vec{x}$. $P(z,t)$ is the prior probability of a specific galaxy is at redshift $z$ and has spectral type $t$, this prior probability can be computed from a spectroscopic sample if one is available. The \pz PDF is, therefore, either the posterior probability, if a prior is used,  or the likelihood itself if no prior is used. This last point arises since the likelihood only depends on the collection of template SEDs; and, if this collection is representative of the overall galaxy sample, the likelihood can be used by itself as a \pz PDF even without a spectroscopic training sample.

\begin{figure}\label{fig:filters}
\includegraphics[width=0.44\textwidth]{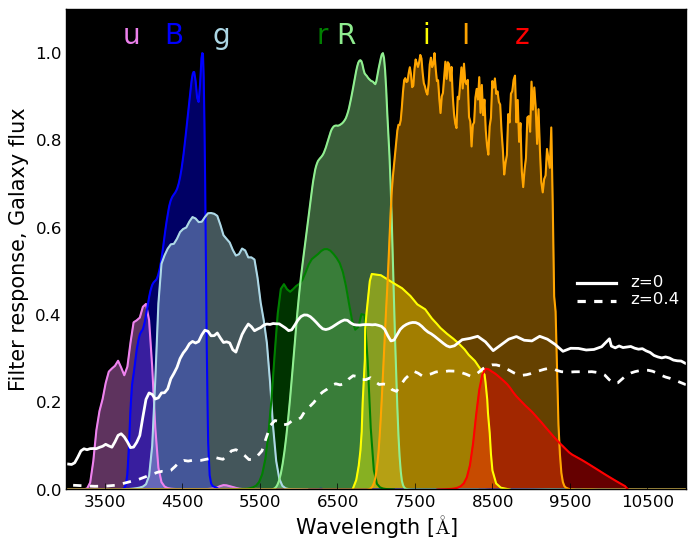}
\caption{ An Elliptical galaxy spectrum at z=0 and redshifted to z = 0.4 overlaid by the eight photometric filters from the DEEP2 galaxy survey (3 from the original survey and $ugriz$ from a matched catalog \citep{Matthews2013}).} 
\label{fig:bpz_example}
\end{figure}

The use of a prior in a Bayesian analysis, however, is recommended. In this case, the prior probability can be computed directly from physical assumptions, from an empirical function calibrated by using a spectroscopic training sample~\citep[\eg][]{Benitez2000}, or from an empirical function calibrated by using machine learning techniques~\citep[see \eg][where we used Random Na\"{\i}ve Bayesian methods to compute the prior probabilities]{CarrascoKind2013b}. For example,~\cite{Benitez2000} propose the following function for a single magnitude $m_0$:
\begin{multline}
 P(z,t\mid m_0) = P(t\mid m_0)P(z\mid t,m_0) \\  
 \propto f_T e^{-k_t (m-m_0)} \times z^{\alpha_t} \exp\left( -\left[\frac{z}{z_{mt}(m)} \right]^{\alpha_t}\right).
\end{multline}
where $z_{mt}(m) = z_0t + k_{mt} (m-m_0)$. The five parameters of this function: $f_T$, $m_0$, $\alpha_t$, $z_{mt}$, and $k_{mt}$ can be constrained either by using direct fitting routines, or by using Markov Chain Monte Carlo methods to sample these parameters. These five parameters are dependent on the template $t$ and can be quantified independently. For additional details on the underlying Bayesian approach, we refer the reader to the original paper by ~\cite{Benitez2000}.

As the goal of a template fitting method is to minimize the difference between observed and theoretical magnitudes (or colors), this approach is heavily dependent on both the library of galaxy SED templates that are used for the computation and the accuracy of the transmission functions for the filters used for  particular survey. SED libraries are generally built from a base set of SED templates. These base templates broadly cover the Elliptical, Spiral, and Irregular categories, and a template library can be constructed by interpolating between the base spectral templates to create new spectra. One of the most widely used set of base templates are the four CWW spectra~\citep{Coleman1980}, which include an Elliptical, an Sba, an Sbb, and an Irregular galaxy template. When extending an analysis to higher redshift, these temples are often augmented with two star bursting galaxy templates published by~\cite{Kinney1996}. One additional effect some template approaches consider is the presence of 
interstellar dust, 
which will introduce artificial reddening.

Once the library of galaxy SED templates has been constructed, the templates are convolved with the transmission functions for a particular survey to generate synthetic magnitudes as a function of redshift for each galaxy template. For the most accurate results, these transmission functions should include the effects of the Earth's atmosphere (if the observations are ground-based), as well as all telescope and instrument effects. This convolution process is demonstrated visually in Figure~ \ref{fig:bpz_example}, which presents an example Elliptical galaxy spectral template at redshift zero and at a redshift 0.4. Overplotted on this figure is the filter set ($B$, $R$, and $I$) used by the DEEP2 survey, which is the data analyzed in this paper, along with the five extra filters: $u, g, r, i, z$ presented in the DEEP2 photometry catalog compiled by \cite{Matthews2013}.

\section{Photo-$Z$ PDF Combination Methods}\label{combine}
We now turn our attention to the different methods with which we can combine distinct \pz PDF estimation techniques~\citep[see \eg][where we first discussed combining Bayesian and machine learning predictions]{CarrascoKind2013b}. In the statistics and machine learning communities, this topic is known as \textit{ensemble learning} \citep{Rokach2010}. Recently, \cite{Dahlen2013} have demonstrated that, on average, an improved \pz estimate can be realized by combining the results from multiple template fitting methods. In this section, we build on this previous work to identify how Bayesian techniques can be used to construct a combined \pz PDF estimator.

We can frame the problem mathematically by writing the set of \pz PDFs  for a given galaxy as a set of models $\vec{M}$,  where each individual model $M_k$ (\eg \tpzns, \somzns, or modified \bpzns) provides a distinct \pz PDF or posterior probability. A \pz PDF can be written as $P(z \mid \vec{x}, \vec{D}, M_k)$, where $\vec{x}$ is the set of magnitudes or colors (note that without loss of generality we can use other attributes in this process) used to make the prediction and $\vec{D}$ corresponds to the training set which consists of $N_d$ galaxies. We can also abbreviate this \pz PDF as $P_k(z)$. These \pz PDFs are each subject to the following constraint:
\begin{equation}\label{pz1}
 \int_{z_1}^{z_2} P_k(z) dz = 1
\end{equation}
for every model $M_k$, where $z_1$ and $z_2$ are the lower and upper limits, respectively, for the redshift range spanned by the galaxy sample. In the following subsections, we introduce different methods to aggregate these \pz PDFs and show the results of these different methods in \S\ref{App}.

Given the variety of \pz PDF estimation methods we are using (\ie supervised, unsupervised, and model-based), we fully expect the relative performance of the individual techniques to vary across the parameter space spanned by the data. For example, supervised methods should perform the best in areas populated by high quality training data, while unsupervised or model-based methods should perform better where we have little or no training data. As a result, we can bin a specific subspace of our multi-dimensional parameter space and apply an individual combination method to each bin separately. This technique is demonstrated later in more detail with the Bayesian Model Averaging method (although it is more generally applicable).

\subsection{Weighted Average}\label{addition}
The simplest approach to combine different \pz PDF techniques is to simply add the individual PDFs and renormalize the sum. In this case the final \pz PDF is given by:
\begin{equation}
 P(z \mid \vec{x},\vec{M})=\sum\limits_{k}P(z \mid \vec{x},M_k) .
\end{equation}
We can improve on this simple approach by including weights in the previous equation:
\begin{equation}
 P(z \mid \vec{x},\vec{M})=\sum\limits_{k} \omega_k P(z \mid \vec{x},M_k) .
\label{swa}
\end{equation}
These weights, $\omega_k$, can be estimated for each input method by using the cross validation or OOB data, or from an intrinsic characteristic of the \pz PDF, such as $zConf$ that we introduced in CB13.  In this work we use three weight schemes in addition to the uniform case:

\subsubsection*{PDF shape weights}
In this case, $\omega_k$ is given by the the $zConf$ parameter, which is similar to the \textit{odds} parameter presented in \cite{Benitez2000} $zConf$ is defined as the integrated probability between $z_{\rm phot} \pm \sigma_{k}(1+z_{\rm phot})$, where $z_{\rm phot}$ is a single estimated value for the \pz PDF. This single \pz estimate can be either the mean or the mode of the \pz PDF. Likewise, we can estimate $\sigma_k$ for each input method either by using the OOB data, by selecting a constant value across all input methods, or by selecting these values separately so that all \pz PDFs have the same cumulative $zConf$ distributions.  $zConf$ quantifies the sharpness of the PDF and can take values from zero to one. In CB13 and CB14, we demonstrated that there is a correlation between this value and the accuracy of the overall \pzns. Specifically, we observed that, on average, galaxies with higher $zConf$ have more accurate \pz PDFs than galaxies with lower $zConf$ values.

\subsubsection*{Best fit weights }
An alternative method to compute the values of $\omega_k$ is to use the cross-validation data to first determine the weight values that minimize the difference between $z_{\rm phot}$ and $z_{\rm spec}$; and, second to apply these best fit values to the test data. This method seeks the optimal linear combination of each individual PDF, thus it allows the values of $\omega_k$ to be negative. After the combination is completed, we renormalize according to Equation \ref{pz1}. This method can be applied to a binned sub-sample to take advantages of the performance of each method in different areas of the attribute space.

\subsubsection*{Oracle scheme}

As mentioned, when the input, multi-dimensional data have been binned (c.f. Figure~\ref{fig:combined_map_cfh}), we can use the cross-validation data to select only one model from among all available input models to only be used with the test data located within that specific bin. Since we are allowed to only select one input model, this will result in an assigned weight value of one for the chosen model and zero otherwise, however the chosen model is allowed to vary between bins.\\

The primary disadvantage of these simple, additive models is that incorrect estimates for the errors for the selected input model can bias the final result. On the one hand, if a technique has underestimated errors, the final result will be biased towards this one input method. On the other hand, overestimation of the errors will bias the final result away from  this particular method. One approach to address this issue, as discussed by~\cite{Dahlen2013}, is to either smooth or sharpen the \pz PDFs estimated by each method by using the OOB data until their error distributions are approximately Gaussian with unit variance. We can generalize this approach to transform a \pz PDF as $P_k(z) = P_k(z)^{\alpha_k}$, where we adjust the value of $\alpha_k$ by using either the cross validation data when errors are over estimated or use a Gaussian smoothing filter when they are under estimated.

\subsection{Bayesian Model Averaging}\label{bma}

Bayesian Model Averaging (BMA) is an ensemble technique that combines different models within a Bayesian framework. BMA accounts for any uncertainty in the correctness of a given model by integrating over the model space and weighting each model by the estimated probability of being the \textit{correct} model. As a result, BMA acts as a model selection procedure that handles the uncertainty in selecting the best model by using a combination of models instead. This is because BMA considers the uncertainty in selecting the best model while working under the assumption that only one model is actually the best~ \citep{Monteith2011}. BMA has been used for astrophysical problems~\citep[see \eg][]{Gregory1992,Trotta2007, Debosscher2007} in, for example, the determination of cosmological parameters and variable star classification~\citep[see,][for a review on using BMA in astronomy]{Parkinson2013}.

When using BMA, the training data are used to characterize each of the models that will be combined. For each galaxy, the final PDF, $P(z \mid \vec{x}, \vec{D}, \vec{M})$, is given by:
\begin{equation}\label{BMA1}
 P(z \mid \vec{x},\vec{D},\vec{M}) = \sum\limits_{k}P(z \mid \vec{x},M_k)P(M_k \mid \vec{D}) .
\end{equation}
$P(M_k \mid \vec{D})$ is the probability of the model $M_k$ given the training data $\vec{D}$, which can be viewed as a simple, model dependent weighting scheme. This probability can be computed by using Bayes' Theorem:
\begin{multline}\label{BMA2}
 P(M_k \mid \vec{D}) = \frac{P(M_k)}{P(\vec{D})} P(\vec{D} \mid M_k) \\
 \propto P(M_k)  \prod\limits_{i=1}^{N_d} P(d_i \mid M_k) .
\end{multline}
We have omitted  the $P(\vec{D})$ term as it is merely a normalization factor and we use the same data for all models. $d_i$ is the $i^{\textrm{th}}$ element from the training data $\vec{D}$, which are assumed to be independent. 

For each model, we assign the value $\epsilon_k$ as an average error for the estimation process. $\epsilon_k$ can be computed as the fraction $N^{(b)}_k/N_d$, where $N^{(b)}_k$ is the number of galaxies considered to be misestimated or \textit{bad} for the particular \pz PDF method $k$. To quantify when a specific galaxy is a bad prediction we compute
\begin{equation}\label{BMA3}
N^{(b)}_{k,i} =
\left\{
\begin{array}{ll}
1  & \mbox{if } \int_{z_s-\delta_z}^{z_s+\delta_z} P(z \mid \vec{x},d_i) dz \leq \pi_z ,\\
0  & \mbox{otherwise} .
\end{array}
\right.
\end{equation}
In this equation, $z_s$ is the spectroscopic redshift for the $i^{\textrm{th}}$ training set galaxy. The first parameter, $\delta_z$, controls the width of a window centered on $z_s$ within which we accumulate \pz probability for the $i^{\textrm{th}}$ training galaxy around the true redshift. The second parameter, $\pi_z$, is the minimum probability within this window for which we consider the model prediction to be  good. We find that $\pi_z = 0.5$ and $\delta_z = 0.05$ provides a good discriminant between good and bad \pz model estimates. 

Given the individual good/bad predictions for each training set galaxy, we can compute the total number of bad predictions, $N^{(b)}_k$, by summing over the individual predictions, $N^{(b)}_{k,i}$, for the entire training data, $\vec{D}$. The total number of good prediction will naturally be $N_d-N^{(b)}_k$. As a result, we can rewrite Equation~\ref{BMA2}:
\begin{equation}
P(M_k \mid \vec{D}) \propto P(M_k) (1-\epsilon_k)^{N_d-N^{(b)}_k}(\epsilon_k)^{N_k^{(b)}},
\end{equation}
where $P(M_k)$ is the probability of each model $k$, which we can assume to be unity for all models. Therefore, the final PDF for each galaxy is given by
 \begin{multline}\label{final_P}
 P(z \mid \vec{x},\vec{D}, \vec{M}) \propto \sum\limits_{k}P(z \mid \vec{x},M_k) P(M_k)  \times \\
 (1-\epsilon_k)^{N_d-N^{(b)}_k}(\epsilon_k)^{N_k^{(b)}} .
\end{multline}

We applied the BMA technique to individual bins within the multi-dimensional parameter space  occupied by a given data set. We demonstrate this binned BMA technique in Figure~\ref{fig:combined_map_cfh}, where we use a Self Organized Map to project our entire input parameter space to a two-dimensional map.  In this manner, all magnitudes or colors are used to form the binned regions within which the parameters of the ensemble learning approach can vary. After computing \pz PDFs for all galaxies with each method, we use BMA to determine the relative weights for these input techniques within each bin; we can visualize these weights as different colors across the two-dimensional map, as shown in Figure~\ref{fig:combined_map_cfh}. This figure graphically displays how the \textit{accuracy} of each \pz PDF estimation varies across the parameter space, and thus how the different weights themselves vary.

\subsection{Bayesian Model Combination}\label{bmc}

As discussed, Bayesian Model Averaging tries to select the best model among the ones introduced to the algorithm. Alternatively, we can modify BMA to produce an more optimal model combination technique~\citep{Monteith2011} known as Bayesian Model Combination (BMC). With BMC, instead of directly combining the three different \pz PDF estimates as was the case with BMA, the Bayesian process is used to explore different combinations of the individual \pz PDF techniques. Thus, an ensemble of different \pz PDF combinations are generated and we directly compare different model combinations. 

As a simple example, we could first generate  hundreds different random weights for all three of our \pz PDF estimation techniques, and second use these to compute hundreds of new \textit{sets} of PDFs by computing a simple weighted average by using Equation~\ref{swa}.  Finally, we could apply BMA to this PDF ensemble to determine the final PDF. In this case, we could write Equation~\ref{BMA1}:
\begin{equation}\label{BMC1}
 P(z \mid \vec{x},\vec{D},\vec{M}, \vec{E}) = \sum\limits_{e \in \vec{E}}P(z \mid \vec{x},\vec{M},e)P(e \mid \vec{D}) ,
\end{equation}
where $e$ is an element from the set $\vec{E}$ of these hundreds combined models. Here we need to compute the performance of each combination $e$ and apply the BMA formulation, shown in Equations~\ref{BMA2} and \ref{BMA3}, to those models by using the model $e$ instead of $M_k$, i.e.,
\begin{equation}
  P(e \mid \vec{D}) \propto P(e) \prod\limits_{i=1}^{N_d} P(d_i \mid e) .
\end{equation}
Fundamentally, with BMC we are marginalizing over the uncertainty in the correct model combination, where in BMA we marginalized over the uncertainty in identifying the correct model from the entire ensemble. 

The number of model combinations $\vec{E}$ is, in principle, infinite, and in practice can be very large. To overcome this, we can use sampling techniques over a reasonable, finite number of models. Naively we might use randomly generated weights, however, this approach can be costly to fully span the allowed range of weights and convergence towards a satisfactory solution might be slow. Thus, instead of assigning weights randomly or using incremental steps within a regular grid, we sample the weights from a Dirichlet distribution where the \textit{concentration} parameters are modified until they converge to stable values. We require that the set of weights, $w_k$, for each of the three models, $M_k$, satisfy $\sum w_k = 1$ and also $w_k > 0$.

For a concentration parameter $\boldsymbol{\alpha}$ of the same dimension as $\vec{w}$, we have that the probability distribution for $\vec{w}$ is given by:
\begin{equation}
P(\vec{w}) \sim \mathcal{D}{\rm ir}(\boldsymbol{\alpha})  = \frac{\Gamma(\sum_k \alpha_k)}{\prod_k \Gamma(\alpha_k)}\prod\limits_k w_k^{\alpha_k -1} ,
\end{equation}
where $\mathcal{D}{\rm ir}(\boldsymbol{\alpha})$ is the \textit{Dirichlet} distribution, $\Gamma(\alpha_k)$ is the \textit{gamma function} and $k$ are the base models, which in this paper are \tpzns, \somzns, and our modified  \bpzns. In order to generate a set $\vec{E}$ of combined models, we first set $\alpha_k$ to unity for all values of $k$. Second, we sample from this distribution $n_s$ times ($n_s$ is a fixed number, generally between 2 and 5, which we fixed at 3) to get  a set of $n_s$ weights and $n_s$ new model combinations. Next, we compute $P(e \mid D)$ by using Equations \ref{BMA2} and \ref{BMA3} for each model in the set of $n_s$ models. We, temporarily,  select the best model among the set $n_s$, i.e, the one with highest $P(e \mid \vec{D})$, and update the $\alpha_k$ parameters by simply adding the weights from the corresponding model to the current values of $\boldsymbol{\alpha}$,
\begin{equation}
\boldsymbol{\alpha}^{(t+1)} = \boldsymbol{\alpha}^{t} + \max_{\vec{w}_e \in n_s} P(e \mid \vec{D}) 
\end{equation}
where $t$ is just a symbolic reference to the fact that $\boldsymbol{\alpha}$ is being updated every 3 steps. 

We use the latest values for $\boldsymbol{\alpha}$ to continue the sampling process to obtain the next set $n_s$  of model combinations. As a result, we continually (by adding $n_s$ new models at each step) extend our set of model combinations $\vec{E}$. As the chain of models in this set is constructed iteratively, the process can be terminated either when a predefined number of model combinations has been reached or when new model combinations have started to converge. This process behaves similarly to a Markov Chain Monte Carlo process, and we have an analogous phase to the \textit{burn in} step, where we can omit some number of model combinations at the start of our set $\vec{E}$ of model combinations. Thus, our final \pz PDF prediction is the application of BMA over the remaining elements in $\vec{E}$, we have set for this work the size of $E$ to be 800. Finally, we note that, as was the case with BMA, we can develop a 
binned version of our BMC technique, where we develop different  model combinations  for different region of 
the magnitude (color) space by using a SOM.

\subsection{Hierarchical Bayes}

A Hierarchical Bayesian (HB) method provides a different approach to combine the individual \pz PDFs. In a manner similar to BMA, we include the uncertainty that a given \pz PDF for a specific galaxy might be incorrectly predicted as a set of nuisance parameters over which we later marginalize. 

Adopting our previous notation, we follow a similar approach to  \cite{Fadely2012} and \cite{Dahlen2013}, and we write the \pz PDF for an individual galaxy for each base method $k$:
\begin{multline}\label{HB1}
 P(z \mid \vec{x},\vec{D},M_k, \theta_k) = \sum\limits_j P(z \mid \vec{x}, \vec{D}, M_k, \theta_{k j}) \times \\
  P(\theta_{k j} \mid \vec{D}, M_k) ,
\end{multline}
where we have introduced the \textit{hyperparameter} $\theta_{k}$, a nuisance parameter that characterizes our uncertainty in the prior distribution of model $k$. The parameter $\theta_{k}$ can be quantified in different forms, but essentially is the misclassification probability of the $k^{\textrm{th}}$ method. Thus, we quantify this mis-prediction probability with $P(\theta_k)$; and we drop the dependence on $\vec{x}$, the measured galaxy attributes, as it does not directly affect the parameter $\theta_k$. Since we will marginalize over $\theta$, we keep the term $\vec{D}$ as we can use the training data to place limits on $\theta_k$ by using the cross-validation data. We note that these probabilities are subject to:
\begin{equation}
 \sum\limits_{j} P(\theta_{k j} \mid \vec{D}, M_k) = 1 .
\end{equation}

If we consider the case where galaxies are predicted correctly or are outliers, $j$ is a binary state. In this model, if we assume that $\gamma_k$ is the fraction of galaxies that are mispredictions or are labeled as outliers for method $k$, we have: $P(\theta_{k 0} \mid \vec{D}, M_k) = \gamma_k$ and $P(\theta_{k 1} \mid \vec{D}, M_k) = (1 - \gamma_k)$. In this case, Equation~\ref{HB1} becomes:
\begin{multline}
P(z \mid \vec{x},\vec{D},M_k, \theta_k) = P_{def}(z \mid M_k, \theta_k)\gamma_k + \\
P(z \mid \vec{x}, \vec{D}, M_k, \theta_k)(1-\gamma_k) ,
\end{multline}
where $P_{def}(z \mid M_k, \theta_k)$  is the default PDF that should be used for the $k^{\textrm{th}}$ method when the original PDF for that method has been determined to be mis-predicted or wrong. In the second term, we use the original PDF for the method $k$, which is multiplied by the fraction of well predicted objects $1-\gamma_k$.

The final PDF after we combine the different \pz PDFs from our base methods in the HB approach is given by:
\begin{equation}
 P(z \mid \vec{x},\vec{D},\theta) = \prod\limits_k  P(z \mid \vec{x},\vec{D},M_k, \theta_k)^{1/\beta} .
\end{equation}
Here, following~\cite{Dahlen2013}, we have introduced an extra parameter $\beta$, which is a constant value that quantifies the degree of covariance between the different base methods. $\beta =1$ corresponds to complete independence between the base methods, while $\beta=3$ (or, more generally, the total number of methods) would correspond to full covariance between them. We can compute $\beta$ from the OOB sample in such way the final error distribution follows a normal distribution with zero mean and unit variance, as we have done in this paper. Alternatively, we can marginalize over all possibles values of $\beta$ when no cross validation data is available and we can integrate over the uncertainty of this parameter.

Finally, by marginalizing over $\theta$ we have our final PDF:  $P(z \mid \vec{x}, \vec{D} )$, or simply $P(z)$ given by:
\begin{equation}
 P(z) = \int_0^1 P(z \mid \vec{x},\vec{D},\theta)P(\theta)d\theta ,
\end{equation}
where $P(\theta)$ is a constant which in the simple case is equal to unity.  If OOB data is available, we can narrow down the range of allowed values for $\theta$ (or effectively $\gamma_k$), so we can set up a limited range for $\gamma_k$ based on the performance of each method $k$ on this data. In this case, $P(\theta)$ will act as a top-hat window function. In any case, the final $P(z)$ is subject to Equation~\ref{pz1}. As discussed before, we can either apply the HB approach to the entire data set, or we can partition the input space and apply the HB approach independently to the binned regions of the parameter space.

\begin{table}
 \begin{minipage}[]{0.48\textwidth}
\caption{The \pz PDF combination methods, their weights and abbreviations presented in  this paper.}
\label{tab:methods}
\renewcommand{\footnoterule}{}
\begin{tabular}{lll}
Method & Weights\footnote{if applicable}  & Abbreviation \\
\hline
Weighted Average & Uniform & ${\rm WA}_{\rm flat}$ \\
Weighted Average & $zConf$ & ${\rm WA}_{\rm shape}$ \\
Weighted Average & best fit& ${\rm WA}_{\rm fit}$ \\
Weighted Average & oracle predictor & ${\rm WA}_{\rm oracle}$ \\
Bayesian Model Averaging & & ${\rm BMA}$\\
Bayesian Model Combination &  & ${\rm BMC}$\\
Hierarchical Bayes &  & ${\rm HB}$\\
\hline
\end{tabular}
\end{minipage}
\end{table}
\section{DATA}\label{deep2data}

To explore different configurations and to demonstrate the capabilities and the efficacy of these  \pz combination techniques, we follow the approach we presented in CB13 and CB14, but in this paper we restrict our analysis to data obtained by the Deep Extragalactic Evolutionary Probe (DEEP) survey and the Sloan Digital Sky Survey (SDSS). In the rest of this section we provide a summary of these  data  and detail how we extracted the data sets from these surveys that we use in the analysis presented in \S\ref{App}.

\subsection{Deep Extragalactic Evolutionary Probe}
 
The DEEP survey is a multi-phase, deep spectroscopic survey that was performed with the Keck telescope. Phase I used the Low Resolution Imaging Spectrometer (LIRS) instrument~\citep{Oke1995}, while phase II used the DEep Imaging Multi-Object Spectrograph (DEIMOS)~\citep{Faber2003}. The
DEEP2 Galaxy Redshift Survey is a magnitude limited spectroscopic survey of objects with $R_{AB} < 24.1$~\citep{Davis2003,Newman2013a}. The survey includes photometry in three bands from the Canada-France-Hawaii Telescope (CFHT) 12K: $B$, $R$, and $I$ and it was recently extended by cross-matching the data to other photometric data sets. In this work, we use  Data Release 4~\citep{Matthews2013}, the latest DEEP2 release that includes secure and accurate spectroscopy for over 38,000 sources. The original input photometry for the sources in this catalog was supplemented by using two $u$, $g$, $r$, $i$, and $z$ surveys: the Canada-France-Hawaii Legacy Survey~\citep[CFHTLS;][]{Gwyn2012}, and the SDSS. For additional details about the photometric extension of the DEEP2 catalog, see~\cite{Matthews2013}. 

To use the DEEP2 data with our implementation, we have selected sources with secure redshifts (ZQUALITY$\geq 3$), which were securely classified as galaxies, have no bad flags, and have full photometry. Even though the filter responses are similar, the $u$, $g$, $r$, $i$, and $z$ photometry originates from two different surveys and are thus not identical. We therefore only present the results from those galaxies that lie within field 1 that have CFHTLS photometry. Furthermore, we  have corrected these observed magnitudes by using the extinction maps from \cite{Schlegel1998}. In the end, this leaves us with a total of 10,210 galaxies each with eight band photometry and redshifts. From this data set, we randomly select 5,000 galaxies for training and hold the remainder out for testing. The computation of \pz PDFs was completed by using the magnitudes in the bands $B$, $R$, $I$, $u$, $g$, $r$, $i$, and $z$ and their corresponding colors $B-R$, $R-I$, $u - g$, $g - r$, $r - i$, and $i - z$, providing a total of 
fourteen dimensions.

\subsection{Sloan Digital Sky Survey}
The Sloan Digital Sky Survey~\citep[SDSS;][]{York2000} phases I, II  and III conducted a photometric survey in the optical bands: $u$, $g$, $r$, $i$, $z$ that covered more than 14,000 square degrees, more then one-quarter of the entire sky. The resultant photometric catalog contains photometry for over $10^8$ galaxies, making the SDSS one of the largest sky surveys ever completed. The SDSS also conducted a spectroscopic survey of targets selected from the SDSS photometric catalog. In this paper, we use a subset of the spectroscopic data contained within the Data Release 10 catalog~\citep[SDSS-DR10]{Ahn2013}, which includes over two million spectra of galaxies and quasars which include those taken as apart as the Baryonic Oscillation Spectroscopic Survey (BOSS) program \citep{Dawson2013}.

Specifically, we selected galaxies by using the online CasJobs website\footnote{http://skyserver.sdss3.org/CasJobs/} and the following query from the DR10 data base:
\begin{verbatim}
SELECT spec.specObjID,
    gal.dered_u, gal.dered_g, gal.dered_r,
    gal.dered_i, gal.dered_z,
    gal.err_u, gal.err_g, gal.err_r,
    gal.err_i, gal.err_z,
    spec.z AS zs
INTO mydb.DR10_spec_clean_phot
FROM SpecObj AS spec 
JOIN Galaxy AS gal
ON spec.specobjid = gal.specobjid, 
    PhotoObj AS phot
WHERE spec.class = `GALAXY' -- Spectroscopic class 
                            -- (GALAXY, QSO, or STAR)
AND gal.objId = phot.ObjID 
AND phot.CLEAN=1            -- Clean photometry flag 
                            -- (1=clean, 0=unclean)
AND spec.zWarning = 0       -- Bitmask of warning
                            -- vaules; 0 means all 
                            -- is well
\end{verbatim}

We also removed some additional bad photometric observations, ensured the redshift values were positive, and compute colors for the final catalog, which contains 1,147,397 galaxies. The spectroscopic data range from $z \approx 0.02$ up to $z \approx 0.8$; the full spectroscopic redshift distribution of these galaxies is shown in the gray shaded histogram presented in Figure \ref{fig:N_z_sdss}. These data are dominated by the Main Galaxy Sample (MGS) at low redshifts, with mean redshift of $z \sim 0.1$, and by luminous red galaxies (LRG) at higher redshifts, with mean redshift of $z \sim 0.5$. 

From this sample, we randomly selected 50,000 galaxies for training and hold the remaining 1,097,397 for testing. This training set corresponds to approximately 4.5\% of the test set. We note that this is a blind test, as the testing data are not used in any way to train or calibrate the algorithms. Of all the measured attributes in the SDSS photometric catalog, we have only used the nine dimensions corresponding to the five galaxy, extinction corrected, model magnitudes and the four colors derived from these five magnitudes: $u$, $g$, $r$, $i$, $z$, $u - g$, $g - r$, $r - i$, and $i - z$. 

\section{results/discussion}\label{App}

We now turn to the actual application of the ensemble learning approaches described in \S\ref{combine} to the data introduced in \S\ref{deep2data}. We present the seven combination methodologies we use in this section in Table \ref{tab:methods}, which also includes an abbreviated name that we will use to refer to a specific technique. We follow a similar approach to CB14 in order to compare different combination methods, and define the bias to be $\Delta z' = |z_{\rm phot}-z_{\rm spec}|/(1+z_{\rm spec})$. We also present the standard metrics we use to compare the performance of the different combination techniques in Table \ref{tab:def_metrics}. As shown in this table, we define five metrics to address the bias and the variance of the results (the first five rows) and we present three values to characterize the outlier fraction. 

We also use the $KS$ metric, which represents the results of a Kolmogorov--Smirnov test that quantifies the likelihood that the predicted \pz distribution and the spectroscopic redshift distribution $N(z)$ are drawn from the same underlying population. This metric provides a single, robust value to compare both distributions that does not depend on how the results are binned by redshift, and it is defined as the maximum distance between both empirical distributions. 

To determine this statistic, we compute the empirical cumulative distribution function (ECDF) for both distributions. For the spectroscopic sample, the ECDF is defined as:
\begin{equation}
F_{\rm spec} (z) =  \sum\limits_{i=1}^{N} \Omega_{z_{\rm spec}^i < z}
\end{equation}
where N is the number of galaxies in the redshift sample, and
\begin{equation}
 \Omega_{z_{\rm spec}^i < z} = \begin{cases} 1, & \mbox{if } z_{{\rm spec},i} < z  \\ 0 , & \mbox{otherwise } \end{cases}
\end{equation}
The ECDF for the \pz distribution is simply the accumulation of the probability presented in the \pz PDF. The summation is carried out over all galaxies in the sample. Given the ECDF for both the \pz and spectroscopic distributions, we compute the KS statistic as:
\begin{equation}
{\rm KS} = \max_z \left(\lvert\lvert F_{\rm phot} (z) - F_{\rm spec} (z)\rvert\rvert \right)
\end{equation}
Thus, as the KS statistic decreases, the two distributions become more similar.

All of the metrics listed in Table \ref{tab:def_metrics} are positive and characterized by the fact that lower metric values indicate a more accurate \pz PDF. In CB14 we defined a new, meta-statistic called  $I$-score (symbolically represented by $I_{\Delta z'}$) that provides a single statistic to simplify the comparison of different \pz techniques. To compute this metric, we first normalize each set of metrics across all different \pz estimation techniques so that we are not biased by different dynamic ranges. Thus, for example, we first compute the mean and standard deviation for  $<\Delta z'>$ for each combination technique, and subsequently rescale all individual $<\Delta z'>$ values so that this set of values has zero mean and unit variance.

We continue this process for all nine statistics listed in Table~\ref{tab:def_metrics}, and compute their weighted sum to obtain the total $I$-score: 
\begin{equation}
I_{\Delta z'} = \sum w_i M_i, 
\end{equation}
where $M_i$ is the rescaled metric and weight value for metric $i$ out of the nine available. For simplicity, we use equal weights in the remainder of this paper (and thus the $I$-score is simply the average of the nine rescaled metrics for each technique). As a result, the \pz method (or parameter configuration) with the lowest $I$-score will be the optimal estimation technique. On the other hand,  if  we were looking for the technique or the specific parameter configuration with, for instance, the lower outlier fraction, we could assign higher weights accordingly to select the best technique. In this way, we can efficiently select the best method or configuration for specific research requirement.

\begin{table}
\caption{The definition of the metrics used to compare different \pz combination methods.}
\label{tab:def_metrics}
\centering
\renewcommand{\footnoterule}{}
\begin{tabular}{ll}
Metric & Meaning\\
\hline
$<\Delta z'>$ & mean of $\Delta z'$\\
$|\Delta z'|_{50}$ & median of $\Delta z'$ \\
$\sigma_{\Delta z'}$ & Standard deviation of $\Delta z'$ \\
$\sigma_{68}$ & Sigma value at which 68\% of $\Delta z'$ is enclosed \\
$\sigma_{\rm MAD}$ & Median absolute deviation = ${\rm median}(||\Delta z' - |\Delta z'|_{50}||)$\\
${\rm KS}$ & Kolmogorov - Smirnov statistic for $N(z)$\\
${\rm out}_{0.1}$ & Fraction of outliers where $\Delta z' > 0.1$\\
${\rm out}_{2\sigma}$ & Fraction of outliers where $|\Delta z' - <\Delta z'> | > 2\sigma_{\Delta z'}$\\
${\rm out}_{3\sigma}$ & Fraction of outliers where $|\Delta z' - <\Delta z'> | > 3\sigma_{\Delta z'}$\\
$I_{\Delta z'}$ &  $I$-score, a weighted combination of all other metrics.\\
\end{tabular}
\end{table}

\subsection{Cross validation data}
In CB13, we introduced OOB data and demonstrated its use as a cross-validation data set that provided error quantification and overall performance similar to what could be expected when applying an algorithm directly to the test data set. When building a tree with \tpz or a map with \somzns, a fraction of the overall training data, usually one-third, is extracted and not used during the tree/map construction process. The resultant tree/map is subsequently applied to this unused data to make a \pz prediction, and this process is repeated for every tree/map. These \pz predications are aggregated for each galaxy to make a \pz PDF; and by construction a galaxy can never be used to train any tree/map that is subsequently used to predict that galaxy's \pzns. Thus, as long as the OOB data remains similar to the final testing data, the OOB data provide results that will be similar to the final test data results and can be used to guide expectations when applied blindly to other data.

\begin{figure}
\includegraphics[width=0.46\textwidth]{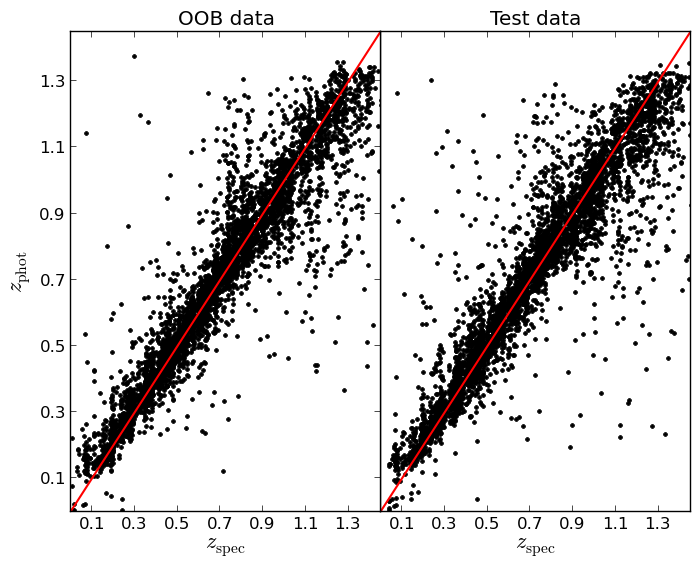}
\caption{A comparison of the photometric (computed by using \somzns) and spectroscopic redshifts for training set (left) and test set (right) galaxies from field 1 of the DEEP2 survey.} 
\label{fig:true_both_deep}
\end{figure}

As an illustration of this process, Figure \ref{fig:true_both_deep} compares the photometric (as computed by using \somzns) and spectroscopic redshifts for galaxies in the training (5,000 in total) and testing (5,210) samples as selected from field 1 of the DEEP2 data set. As shown in this Figure, the performance on both the OOB and the testing data are visually similar and there is no indication of overfitting. In addition, general features in the result, like the spread of the data or the slight tilt of the distribution of points relative to the diagonal, are observed in both samples.

\begin{figure}
\includegraphics[width=0.46\textwidth]{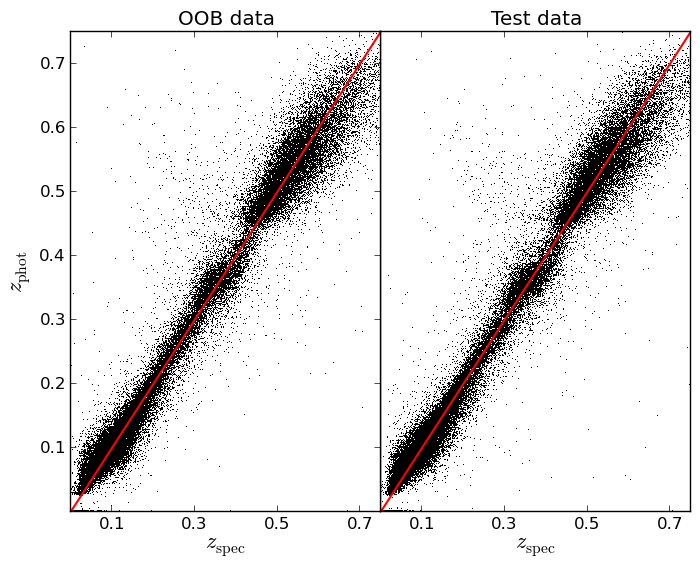}
\caption{A comparison of the photometric (computed by using \tpzns)  and the spectroscopic redshift from the SDSS-DR10 for the 50,000 training set galaxies (left) and 50,000 galaxies randomly subsampled from the 1,097,397 galaxies in the test set (right). } 
\label{fig:true_both_sdss}
\end{figure}

A similar conclusion is observed with the SDSS data, as shown in Figure \ref{fig:true_both_sdss} where the photometric (as computed by using \tpzns) and spectroscopic redshifts for 50,000 galaxies from the training set are compared to 50,000 randomly selected galaxies from the test set. Both distributions show similar behavior and global trends, thus we conclude that, as expected, the OOB data can be used to predict the performance of an PDF combination algorithm on real data.

Another method to contrast the results from these data is to compute the correlation between each of the three \pz estimation techniques discussed earlier as a function of redshift. For this, we use the \pz PDFs for all galaxies, and we calculate the Pearson correlation coefficient $R_{ik}$ within each redshift bin. Even if the three input methods are completely independent, we should expect a positive correlation between them if their predictions are similar. In fact, we desire a positive correlation (but not necessarily a perfect correlation) between the techniques as this will indicate the different techniques are all performing well.

\begin{figure}
\includegraphics[width=0.46\textwidth]{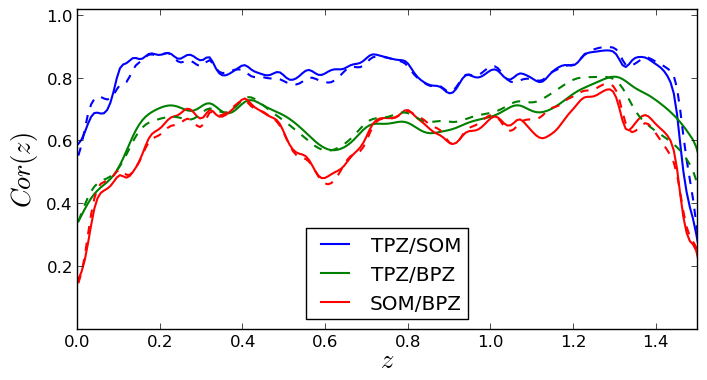}
\includegraphics[width=0.46\textwidth]{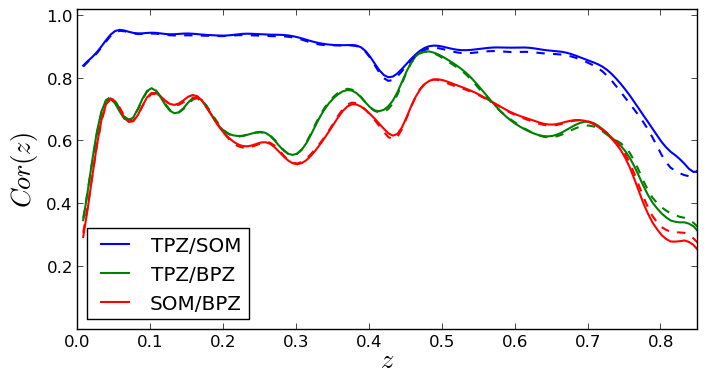}
\caption{The Pearson correlation coefficient between the individual \pz PDF estimation methods as a function of redshift for the DEEP2 (top) and SDSS (bottom) data. The coefficients measured from the cross-validation (OOB) data (dashed line) and from the test data (solid line) are nearly identical, indicating the utility of the OOB data in predicting the performance of an algorithm on blind test data. Note that a positive correlation is beneficial since this measures the relative performance of different techniques in predicting redshifts.}
\label{fig:correlation}
\end{figure}

We present the Pearson correlation coefficient for the three \pz PDF estimation techniques for the DEEP2 data (top panel) and the SDSS data (bottom panel) in Figure~\ref{fig:correlation}. In this figure we display these correlation coefficient computed from the cross-validation (OOB) data (dashed line) and the test data (solid line). The global agreement between these lines further demonstrates the importance of the OOB data as a predictor of the performance of a given technique. This figure also demonstrates a tighter correlation between the two machine learning algorithms than between any machine learning algorithm and the template technique, which is not surprising given the similarities in the methods. While not shown, the shape of the covariance matrices resemble the spectroscopic $N(z)$ distributions presented in Figures \ref{fig:N_z_deep} and \ref{fig:N_z_sdss}. We conclude that this is expected since a larger number of galaxies can naturally produce a greater chance for divergent \pz estimates.

\begin{figure}
\includegraphics[width=0.44\textwidth]{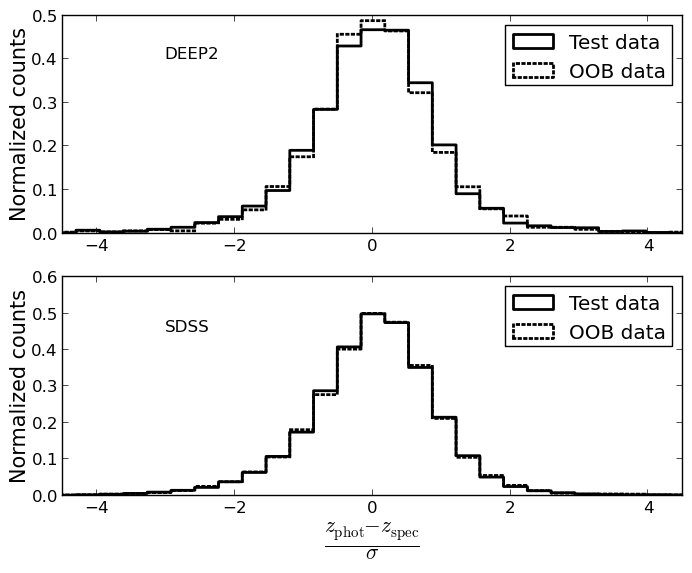}
\caption{The normalized error distributions for galaxies in DEEP2 (top) and SDSS (bottom). The error distribution computed from the test data is shown in red, while the error distribution for the cross-validation (OOB data) is shown in black. The excellent agreement highlights the importance of the OOB data in predicting the results of blind test data predictions.} 
\label{fig:err_test_oob}
\end{figure}

As mentioned previously, a concern when combining \pz PDFs from different methods is to reduce the likelihood of being biased by methods that might under- or overestimate their errors. To further demonstrate the importance of the cross-validation data, we compare the normalized error distribution between the cross-validation (OOB) and test data in Figure \ref{fig:err_test_oob} for both DEEP2 (top panel) and SDSS (bottom panel) data, where the \pz PDFs were generated by \tpz. In both cases, the two curves are nearly identical, and we confirmed the same result with both \somz and \bpzns. Thus we can use the OOB data error estimate to rescale the PDF for the test data by using the results computed from the OOB data.

\subsection{\PZ PDF Combination for DEEP2}

To combine the three \pz PDF techniques discussed in \S \ref{pz_methods}, we employ a binning strategy to allow different method combinations to be used in different parts of parameter space. We first create a two dimensional, $10 \times 10$ SOM representation of the full 14-dimensional space (eight magnitudes and six colors, note that we do not compute a color between the two different photometric input surveys) by using a rectangular topology to facilitate visualization. With this map we can perform an analysis of all galaxies that lie within the same cell, in a similar process to that described in CB14, but now instead of predicting a \pzns, we are computing the optimal model combination. We apply all seven combination methods presented in Table \ref{tab:methods} to all galaxies within each cell by using the OOB data that are also contained within the same cell. We note that the ${\rm WA}_{\rm flat}$ and ${\rm WA}_{\rm shape}$ methods do not depend on this binning, and can, 
therefore, be used without OOB data. We also could employ the ${\rm HB}$ approach without using this map, but in this case we would need to define  $P_{def}(z \mid M_k, \theta_k)$ and perform the marginalization over the entire range of $\theta_k$ without any prior on this value.

\begin{figure}
\includegraphics[width=0.44\textwidth]{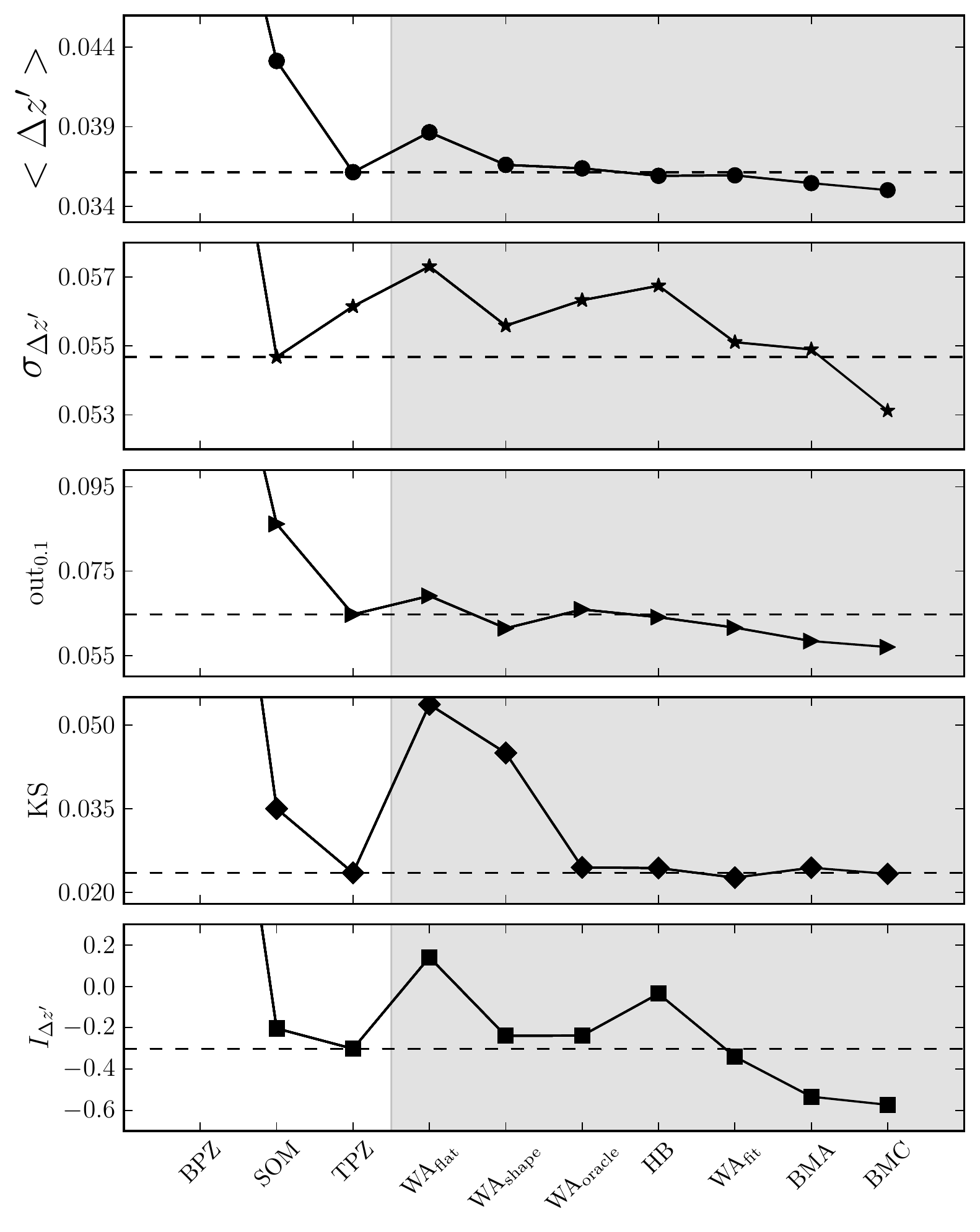}
\caption{A comparison of the average performance for the three individual \pz PDF estimation methods and the seven different \pz PDF combination approaches for five different metrics as defined in Table \ref{tab:def_metrics} for the DEEP2 data. The horizontal dashed line indicates the best result for a given statistic among the three individual methods (note, \bpz is not always shown at the provided scale), and the shaded area separates the individual methods from the combined approaches. All values are presented in Table \ref{tab:big_metrics_cfh}.} 
\label{fig:metrics_cfh_nocut}
\end{figure}

\begin{table*}
\begin{minipage}[]{\textwidth}
\caption{A summary of the performance results for the three individual methods and the seven different \pz PDF combination  methods as applied to the DEEP2 data, no magnitude cut was applied during the training phase. The bold entries highlight the best value within each column to aid in the interpretation of the table (c.f. Figure~\ref{fig:metrics_cfh_nocut}).}
\label{tab:big_metrics_cfh}
\centering
\renewcommand{\footnoterule}{}
\begin{tabular}{l  c c c c c c c c c c}
\hline
Combination method & $<\Delta z'>$ & $|\Delta z'|_{50}$ & $\sigma_{\Delta z'}$ & $\sigma_{68}$ & $\sigma_{\rm MAD}$ & ${\rm KS}$ & ${\rm out}_{0.1}$ & ${\rm out}_{2\sigma}$& ${\rm out}_{3\sigma}$ & $I_{\Delta z'}$\\
 \hline  \hline 
${\rm TPZ}$ & 0.0361 &0.0205 &0.0561 &0.0257 &0.0139 &0.0235 &0.0647 &0.0307 &0.0184 &-0.3021 \\
${\rm SOM}$ & 0.0431 &0.0291 &0.0547 &0.0325 &0.0188 &0.0350 &0.0862 &\textbf{0.0284}& \textbf{0.0150}& -0.2035 \\
${\rm BPZ}$ & 0.0635 &0.0476 &0.0679 &0.0428 &0.0273 &0.1342 &0.1636 &0.0338 &0.0170 &2.3255 \\
${\rm WA_{\rm flat}}$ & 0.0386 &0.0231 &0.0573 &0.0285 &0.0155 &0.0537 &0.0691 &0.0313 &0.0192 &0.1409 \\
${\rm WA_{\rm oracle}}$ & 0.0364 &0.0206 &0.0563 &0.0260 &0.0139 &0.0245 &0.0659 &0.0313 &0.0184 &-0.2385 \\
${\rm WA_{\rm shape}}$ & 0.0366 &0.0217 &0.0556 &0.0268 &0.0146 &0.0450 &0.0614 &0.0297 &0.0186 &-0.2392 \\
${\rm WA_{\rm fit}}$ & 0.0359 &0.0208 &0.0551 &\textbf{0.0253}& \textbf{0.0137}& \textbf{0.0227}& 0.0616 &0.0318 &0.0178 &-0.3404 \\
${\rm BMA}$ & 0.0355 &0.0211 &0.0549 &0.0257 &0.0140 &0.0245 &0.0584 &0.0289 &0.0178 &-0.5339 \\
${\rm BMC}$ & \textbf{0.0350}& 0.0208 &\textbf{0.0531}& 0.0255 &0.0140 &0.0233 &\textbf{0.0570}& 0.0297 &0.0176 &\textbf{-0.5734} \\
${\rm HB}$ & 0.0359 &\textbf{0.0199}& 0.0568 &0.0259 &\textbf{0.0137}& 0.0244 &0.0641 &0.0329 &0.0196 &-0.0354 \\
 \hline 
\end{tabular}
\end{minipage}
\end{table*}

We present a summary of the results obtained by applying the seven different combination techniques to all the galaxies within the DEEP2 data in Table \ref{tab:big_metrics_cfh}. The bold entries in this Table highlight the best technique for any particular metric. The first three rows in this Table show the individual \pz PDF estimation techniques, of which \tpz generally performs the best and is thus shown in the first row as the benchmark. This Table also clearly indicates that the seven different combination techniques generally have a similar performance, and, as shown in the last four rows, often perform better than \tpzns.

We observe that the last four methods: ${\rm WA_{\rm fit}}$, ${\rm BMA}$, ${\rm BMC}$,  and ${\rm HB}$ all use the binned model combination approach, and thus can take advantage of the different performance characteristics of individual codes. In this case, ${\rm BMC}$ provides the best performance as measured by the $I$-score $I_{\Delta z'}$, the bias $<\Delta z'>$, the scatter $\sigma_{\Delta z'}$, and the outlier fraction ${\rm out}_{0.1}$. Overall, the differences are close to 5\% for many of the metrics, which, while small, are still significant since these are averaged metrics over the full test galaxy sample.

In Figure \ref{fig:metrics_cfh_nocut}, we present a visual comparison between the ten different \pz estimation techniques for five different metrics: bias, scatter, outlier fraction, KS test, and the $I$-score. In each panel, the horizontal dashed line shows the best value from the individual \pz PDF estimation methods and the shaded area separates the individual from the combined methods. This Figure demonstrates that the Bayesian modeling techniques provide better performance than the best individual method over all five metrics, and also that by employing the binning scheme to optimize the combination approach we achieve better performance than for the best individual technique. 

\begin{figure}
\includegraphics[width=0.44\textwidth]{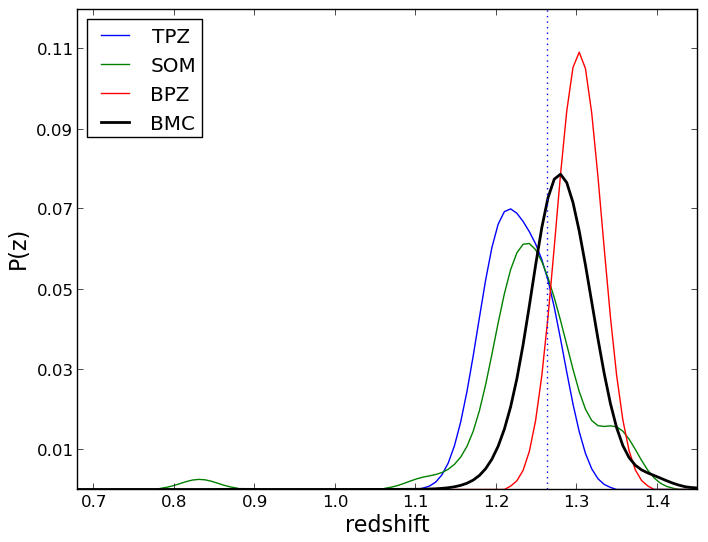}
\caption{An comparison between the three individual \pz PDF estimation techniques and a combined PDF computed by using ${\rm BMC}$ and Equation \ref{final_P} for a single example galaxy taken from the DEEP2. The vertical line indicates the true source redshift.} 
\label{fig:combined_pdf_cfh}
\end{figure}

We compare the actual \pz PDF for a single galaxy selected from the DEEP2 survey as estimated by the three individual techniques with the \pz PDF estimated by the ${\rm BMC}$ method in Figure~\ref{fig:combined_pdf_cfh}. This Figure clearly shows how the re-normalized combined PDF from the three individual \pz PDF estimation techniques has been improved as the ${\rm BMC}$ result is closer to the true galaxy redshift, shown by the vertical line. These combination techniques identify which individual method works best in different cells, and can use that information to either weight the individual \pz PDFs accordingly, or in the case of ${\rm BMC}$ to marginalize over the uncertainty in the correct weights to produce the best combination.

\begin{figure}
\includegraphics[width=0.48\textwidth]{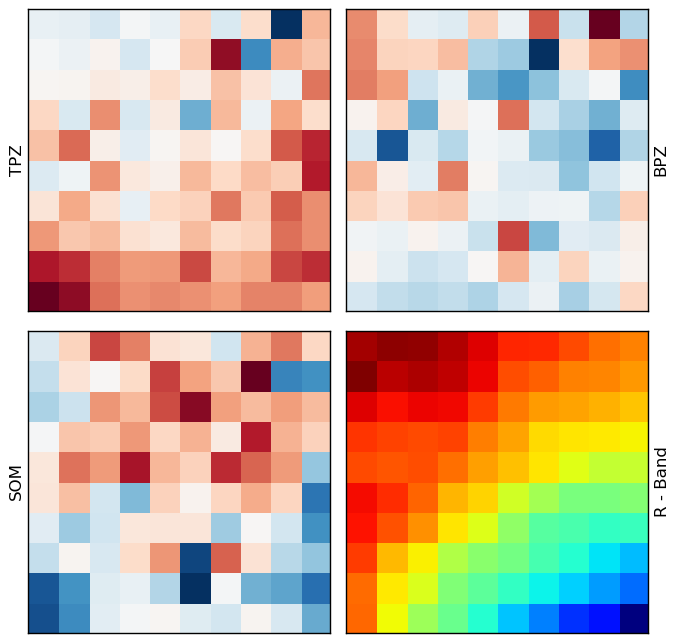}
\includegraphics[width=0.48\textwidth]{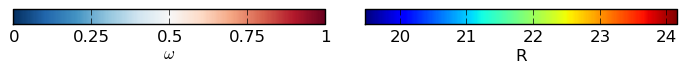}
\caption{A two-dimensional SOM showing the relative weights for the BMA combination scheme applied to the three individual methods for the DEEP2 field 1 data (\tpz is top left, \bpz is top right, and \somz is bottom left). In each panel, the color map indicates the value of the weight relative to the other cells in the map. The bottom right panel shows the same cells colored by the  mean $R$-band magnitude for the cross validation galaxies.} 
\label{fig:combined_map_cfh}
\end{figure}

We apply a SOM to the DEEP2 field 1 data in order to construct a two-dimensional, binned combination of the three individual \pz PDF estimation methods. We use this SOM to determine the weights for the three individual methods for each cell, and present the results in Figure \ref{fig:combined_map_cfh} when using the BMA approach as it is easy to interpret. We also show the mean DEEP2 $R$-band magnitude for all galaxies in a given cell in the lower right panel, which clearly indicates the ability of the SOM to preserve relationships between galaxies when projecting from the higher dimensional space to the two-dimensional map. Of course, the SOM mapping is a non-linear representation of all magnitudes and colors, thus the DEEP2 $R$-band map should only be used to provide guidance.

In the three weight maps, a redder color indicates a higher weight, or equivalently that the corresponding method performs better in that region. These weight maps demonstrate the variation in the performance of the individual techniques across the two-dimensional parameter space defined by the SOM. For example, \bpz performs the best, as expected, in the upper left corner of the map, which is approximately where the faintest galaxies, at least in the DEEP2 $R$-band, are stored. On the other hand, \tpz performs better in the lower sections of the map, which approximates to brighter DEEP2 $R$-band magnitudes. Interestingly, \somz performs relatively better in the upper middle of the map, which corresponds to the middle range $21 \la R \la 23$. The overall variation in weights across the map reflects the performance differences between the individual methods, which are exploited by the combination algorithms in order to identify the optimal combined performance.

\begin{figure*}
\includegraphics[width=0.95\textwidth]{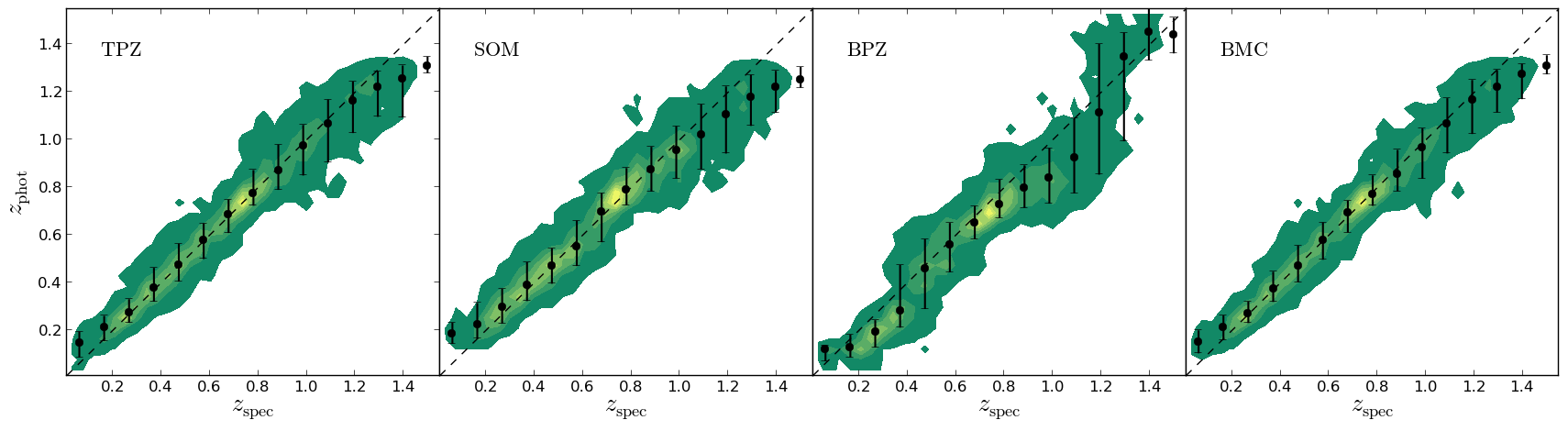}
\caption{A comparison of the photometric and the spectroscopic redshifts for all DEEP2 field1 galaxies. From left to right, the comparison is for the \tpzns, \somzns, \bpzns, and the $BMC$ techniques.The black dots are the median values of $z_{\rm phot}$ and the errors bars correspond to the tenth and ninetieth percentiles within a given spectroscopic redshift bin of width $\Delta z = 0.1$} 
\label{fig:combined_true_deep}
\end{figure*}

\begin{figure}
\includegraphics[width=0.48\textwidth]{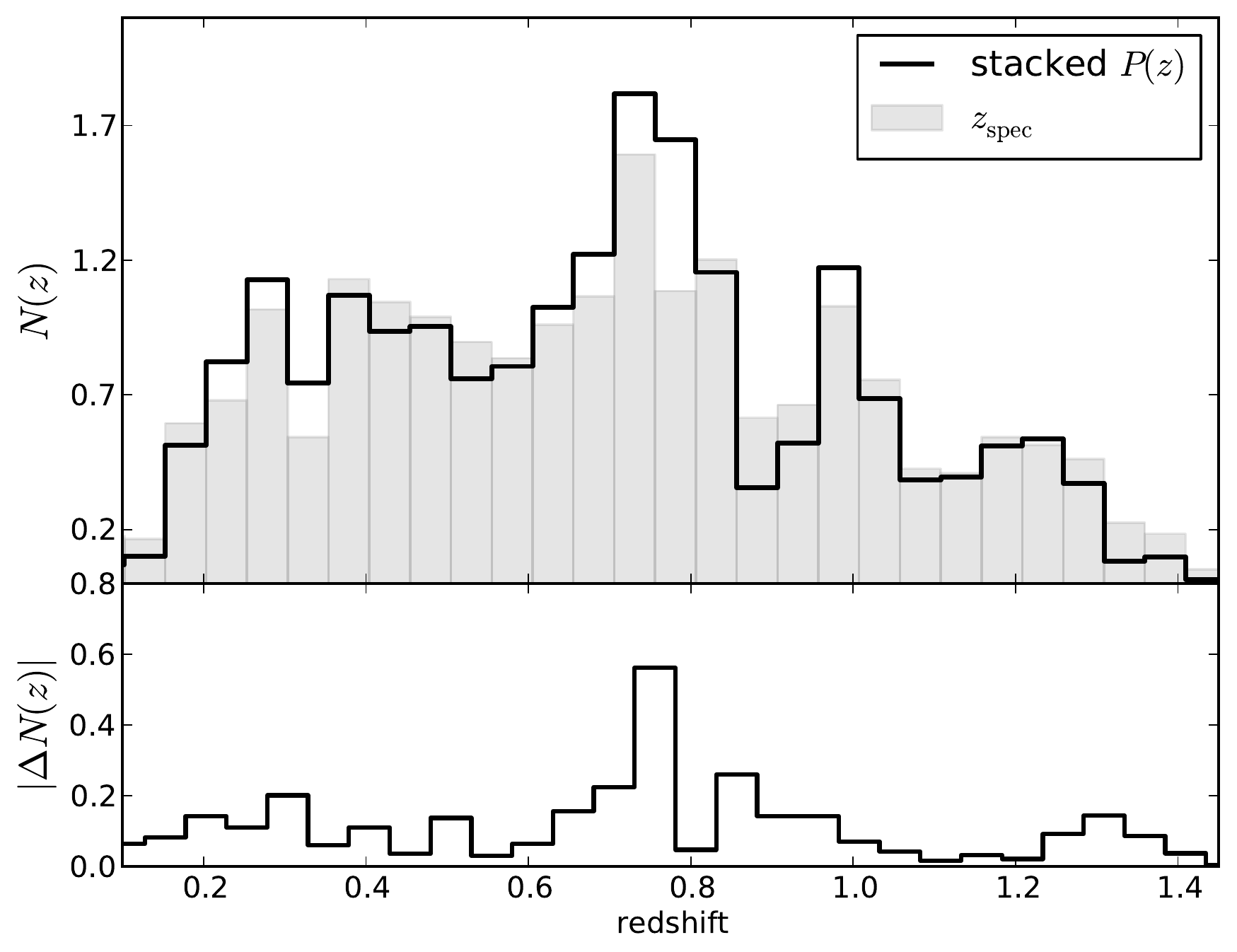}
\caption{Top panel: The $N(z)$ for the DEEP2 sample computed directly from the spectroscopic redshifts (gray) and by stacking the \pz PDF estimates from the $BMC$ method (black). Bottom Panel: The absolute difference between these two $N(z)$ distributions.} 
\label{fig:N_z_deep}
\end{figure}

We can also compare the global performance of the ${\rm BMC}$ method with the three individual \pz PDF methods as a function of the spectroscopic redshift as shown in Figure~\ref{fig:combined_true_deep}. In this Figure, the photometric redshifts are the computed as the mean of each PDF, and the median is shown as black points along with the tenth and ninetieth percentiles as vertical error bars, enclosing 80\% of the distribution on each redshift bin. The performance of the ${\rm BMC}$ method is generally more accurate, resulting in a tighter distribution that suffers fewer outliers when compared to the benchmark \tpz method. Interestingly, the \somz performance is similar to \tpzns, while \bpz is worse, with  wider spread and several discontinuities. Nevertheless, the combined method still uses \bpzns, as shown in the weight maps, as appropriate to generate an overall improved performance, especially for the faintest galaxies as discussed previously. We note, however, that the number counts in the last few 
bins are very low for the DEEP2 training and testing sets 
as shown in Figure~\ref{fig:N_z_deep}. Therefore, although on average \bpz has better performance statistics over those bins (with large error bars), the \pz results remain subject to Poissonian fluctuations (which is important when constructing a SOM to subdivide the galaxies when applying the combination models), thus the BMC results do not emphasize the \bpz  results in the highest redshift bins. 

Of all of the ten different metrics presented in Table~\ref{tab:big_metrics_cfh}, only the $KS$ test does not show a marked improvement over the benchmark \tpz method. This metric does not depend on the redshift binning and it is computed by using the stacked PDF for each method. As a result, this metric is expected to be less sensitive to a combination approach, since stacking the PDF smooths out little discrepancies between the models. After integrating over a large number of galaxies PDFs, the individual methods will not differ significantly from one another and the final $N(z)$ distribution will resemble the one from the benchmark method. 

Figure \ref{fig:N_z_deep} shows the final $N(z)$ produced by stacking the PDFs from the ${\rm BMC}$ technique for galaxies from the DEEP2 (in solid black) and the corresponding DEEP2 spectroscopic $N(z)$ for the same galaxies (in gray). As also seen in CB13 and CB14 for \tpz and \somz respectively, both distributions match exceedingly well.

\subsection{\PZ PDF Combination for the SDSS}

We now change our focus to the analysis of the SDSS galaxy sample, which consists of 1,097,397 galaxies taken from the SDSS-DR10 data; we now retain 50,000 galaxies for training purposes. We apply the same three \pz PDF estimation methods and  seven different combination methods. We construct a SOM-defined, $10 \times 10$ two-dimensional map to subdivide the multi-dimensional magnitude and color space by using a rectangular topology to facilitate visualization. As before, we use cross-validation data to identify the best set of model parameters within each individual cell in our two-dimensional map. As shown in Figures~\ref{fig:correlation} and  \ref{fig:err_test_oob}, the \pz PDFs computed by using the cross-validation and testing data sets are comparable and unbiased. 

We present in Table~\ref{tab:big_metrics_sdss} the same ten metrics for each method, and in bold we highlight the best method for each metric. Overall, the results obtained for this data set are remarkable, especially for the outlier fraction  and the dispersion. We once again treat \tpz as the benchmark method; but note that, interestingly enough,  in two cases, including the $KS$ metric, \tpz does  provide the best result. In addition, both ${\rm BMA}$ and ${\rm BMC}$ have very similar results, with the latter being slightly better. 

After these two models, ${\rm WA_{\rm shape}}$, which is OOB data independent, shows good performance, especially when looking at the $I_{\Delta z'}$ score. For any given individual metric, however, it does not perform better than other combination methods. For this data, \bpz provides good results; thus we expect that the set of template described in \S\ref{template} are a good representation of the galaxies in the SDSS photometric data. In particular, this seems true of the LRGs that dominate this sample for $z \ga 0.3$. 

\begin{table*}
\begin{minipage}[]{\textwidth}
\caption{A summary of the performance results for the three individual methods and the seven different \pz PDF combination methods as applied to the SDSS-DR10 data, with no magnitude cut applied to the training data set. The bold entries highlight the best value within each column to aid in the interpretation of the table (c.f. Figure~\ref{fig:metrics_sdss}).}
\label{tab:big_metrics_sdss}
\centering
\renewcommand{\footnoterule}{}
\begin{tabular}{l  c c c c c c c c c c}
\hline
Combination method & $<\Delta z'>$ & $|\Delta z'|_{50}$ & $\sigma_{\Delta z'}$ & $\sigma_{68}$ & $\sigma_{\rm MAD}$ & ${\rm KS}$ & ${\rm out}_{0.1}$ & ${\rm out}_{2\sigma}$& ${\rm out}_{3\sigma}$ & $I_{\Delta z'}$\\
 \hline  \hline 
${\rm TPZ}$ & 0.0188 &0.0137 &0.0219 &0.0139 &\textbf{0.0082}& \textbf{0.0260}& 0.0078 &0.0297 &0.0121 &-0.2875 \\
${\rm SOM}$ & 0.0201 &0.0149 &0.0209 &0.0152 &0.0094 &0.0381 &0.0070 &0.0334 &0.0125 &0.7836 \\
${\rm BPZ}$ & 0.0230 &0.0164 &0.0289 &0.0167 &0.0103 &0.0367 &0.0134 &\textbf{0.0228}& 0.0111 &1.7143 \\
${\rm WA_{\rm flat}}$ & 0.0195 &0.0139 &0.0235 &0.0145 &0.0088 &0.0292 &0.0082 &0.0251 &0.0104 &-0.2507 \\
${\rm WA_{\rm oracle}}$ & 0.0193 &0.0141 &0.0220 &0.0145 &0.0089 &0.0373 &0.0067 &0.0266 &\textbf{0.0100}& -0.1495 \\
${\rm WA_{\rm shape}}$ & 0.0192 &0.0136 &0.0236 &0.0143 &0.0086 &0.0297 &0.0081 &0.0243 &0.0102 &-0.4114 \\
${\rm WA_{\rm fit}}$ & 0.0200 &0.0141 &0.0242 &0.0149 &0.0090 &0.0274 &0.0090 &0.0255 &0.0107 &0.0244 \\
${\rm BMA}$ & 0.0183 &0.0133 &0.0209 &0.0139 &0.0084 &0.0261 &0.0060 &0.0296 &0.0110 &-0.6384 \\
${\rm BMC}$ & \textbf{0.0183}& \textbf{0.0133}& \textbf{0.0203}& \textbf{0.0138}& 0.0084 &0.0267 &\textbf{0.0059}& 0.0296 &0.0109 &\textbf{-0.6873} \\
${\rm HB}$ & 0.0198 &0.0143 &0.0237 &0.0147 &0.0090 &0.0271 &0.0084 &0.0251 &0.0106 &-0.0975 \\
 \hline 
\end{tabular}
\end{minipage}
\end{table*}

\begin{figure}
\includegraphics[width=0.44\textwidth]{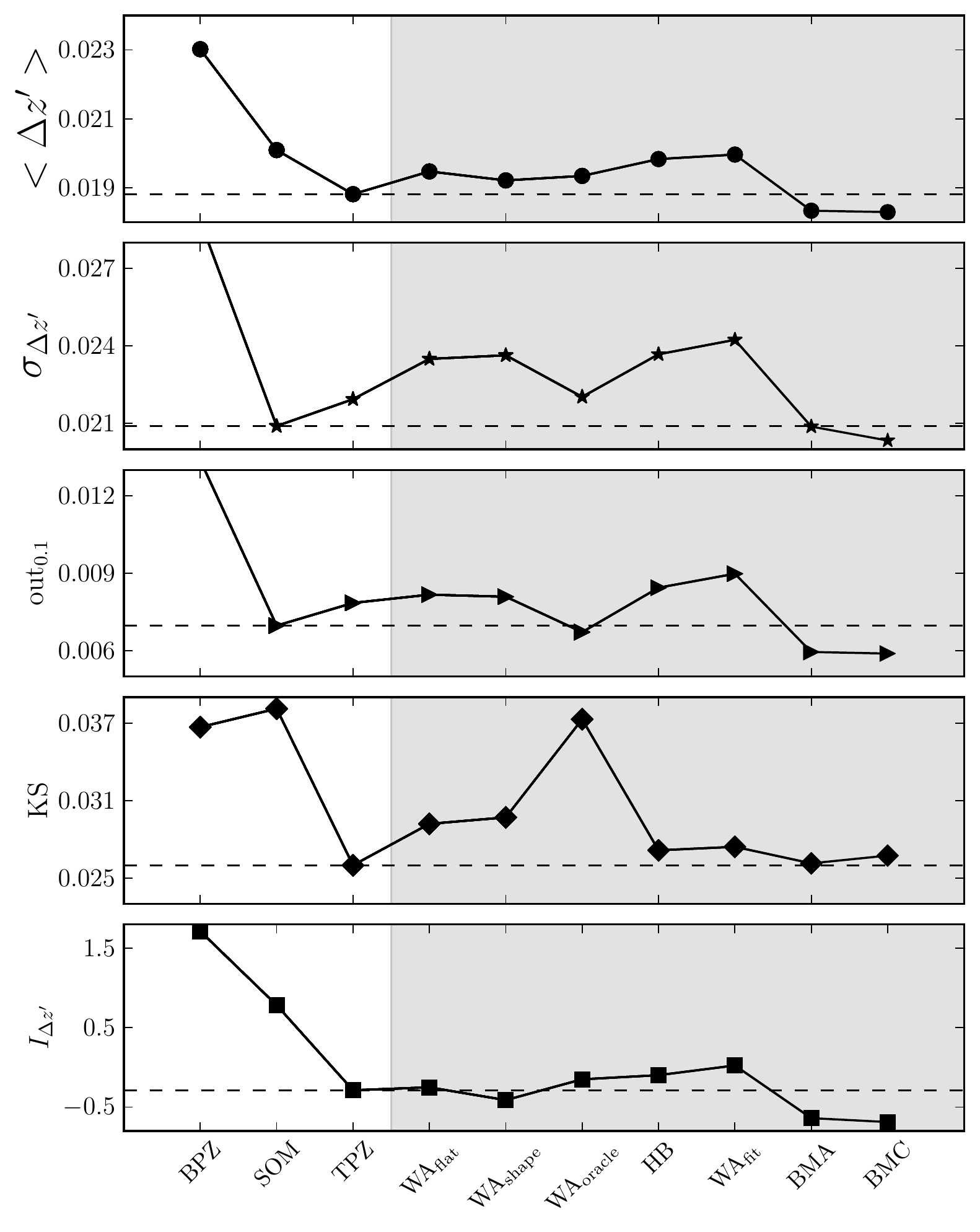}
\caption{A comparison of the average performance for the three individual \pz PDF estimation methods and the seven different \pz PDF combination approaches for five different metrics as defined in Table \ref{tab:def_metrics} for the SDSS data. The horizontal dashed line indicates the best result for a given statistic among the three individual methods, and the shaded area separates the individual methods from the combined approaches. All values are presented in Table~\ref{tab:big_metrics_sdss}. } 
\label{fig:metrics_sdss}
\end{figure}

We present the performance of the three individual and seven combination methods when applied to the SDSS data for five of the most common metrics  in Figure~\ref{fig:metrics_sdss}. As was the case with the DEEP2 data, the Bayesian combination methods provide good performance. We also see the same variation in the $KS$ metric, especially when comparing the combination methods to \tpzns. However, \tpz is not always the best performer among the individual techniques, for example \somz displays the best performance as measured by $\sigma_{\Delta z'}$ and ${\rm out}_{0.1}$. 

As we discussed in CB14, \somz performs quite well when using a spherical topology; in the current application to the SDSS data, we have used a random atlas containing 300 maps that use spherical topology each with 3072 total cells.  Interestingly, the ${\rm WA_{\rm oracle}}$ method, which selects the best method within each binned cell, often selects the \somz result as we can infer from Figure~\ref{fig:metrics_sdss}. Although in general the \textit{oracle} combination method is not the best possible combination, as shown by the overall performance of the ${\rm BMA}$ and ${\rm BMC}$ combination methods on this data.

\begin{figure}
\includegraphics[width=0.48\textwidth]{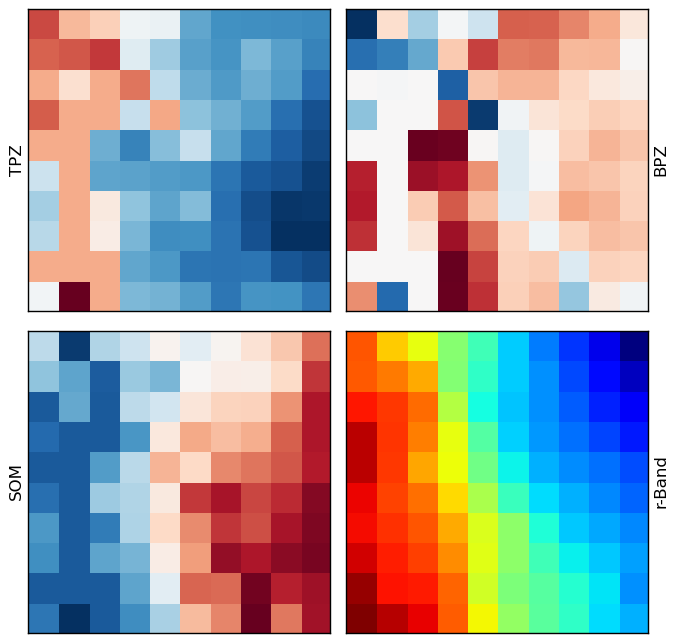}
\includegraphics[width=0.48\textwidth]{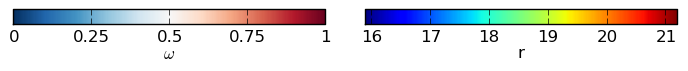}
\caption{A two-dimensional SOM showing the relative weights for the BMA combination scheme applied to the three individual methods for the SDSS data (\tpz is top left, \bpz is top right, and \somz is bottom left). In each panel, the color map indicates the value of the weight relative to the other cells in the map. The bottom right panel shows the same cells colored by the  mean SDSS $r$-band magnitude for the cross validation galaxies.} 
\label{fig:combined_map_sdss}
\end{figure}

We also display the SOM-defined, $10 \times 10$ two-dimensional map used to determine the weights for the three individual methods for each cell in Figure~\ref{fig:combined_map_sdss}. In this map, we identify galaxies within the OOB and test data to determine the parameters for the combination models. One of the benefits of using an unsupervised learning method for this mapping is that we can use any property from the galaxies within this map to construct a representation, such as the mean SDSS $r$-band magnitude map shown in the bottom right panel of Figure~\ref{fig:combined_map_sdss}. In this panel the brighter galaxies are generally on the right while the fainter galaxies are on the left, even though all five magnitudes and four colors were used to construct the SOM-defined, two-dimensional map.

The weighting for the three individual methods show interesting patterns, and \tpz and \somz seem complimentary in that \tpz is weighted most strongly at fainter $r$-band magnitudes (the left side of the map) while \somz is weighted most strongly at brighter $r$-band magnitudes (the right side of the map). This result is most likely an artifact from the bi-modality of the training data, which is dominated at low redshift by the SDSS main galaxy sample and at high redshifts by the SDSS-III LRG sample. At brighter magnitudes and lower redshifts, the \somz approach where a high-dimensional space is projected to two-dimensions does a better job of maintaining complex relationships within the data. At fainter magnitudes and higher redshifts, however, the data are dominated by the homogeneous LRG sample. The \tpz approach performs better for this sample, since the high-dimensional space is recursively sub-divided by \tpz to maximize the information gain, which may only require one or two dimensions.

\begin{figure*}
\includegraphics[width=0.95\textwidth]{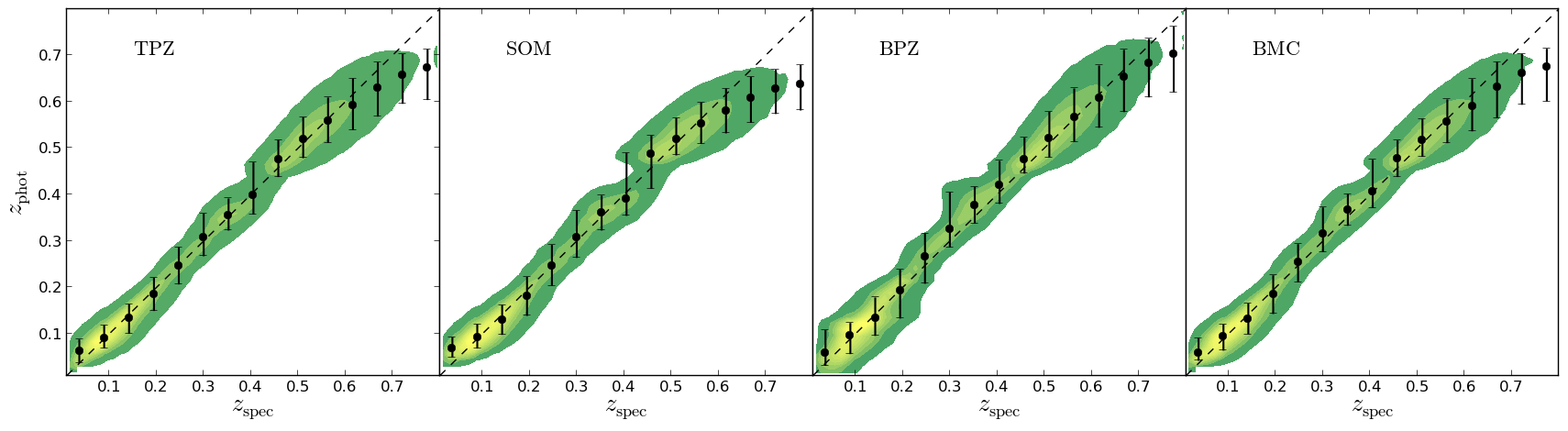}
\caption{A comparison of the photometric and the spectroscopic redshifts for all SDSS galaxies. From left to right, the comparison is for the \tpzns, \somzns, \bpzns, and the $BMC$ techniques.The black dots are the median values of $z_{\rm phot}$ and the errors bars correspond to the tenth and ninetieth percentiles within a given spectroscopic redshift bin of width $\Delta z = 0.05$} 
\label{fig:combined_true_sdss}
\end{figure*}

Another interesting observation from these weight maps is that \bpz performs well over much of the parameter space, with a particular strong weighting in a narrow vertical band on the extreme left of the map and again in the center of the map. Given the nature of the input galaxy sample, it seems reasonable to expect that these areas of the map are dominated by Elliptical galaxies. Another interesting observation is that there are six cells in the second column from the left that all have the same value in each weight map (pink for \tpzns, white for \bpzns, and light blue for \somzns). These cells are primarily empty, \ie they contain weights and training data but they lack test galaxies and thus have a constant value, which illustrates how strongly the galaxies (\ie MGS or LRG) are concentrated in this SOM-defined, two-dimensional topology. 

The number of galaxies, either for training or testing, within each cell can vary significantly, which is simply due to the fact that we used a fixed number of cells (in this case 100) to represent the higher dimensional space when fewer cells would have been sufficient. However, the empty cells do not affect the performance of the \pz combination methods, they are simply not used during the analysis. It is the fact that these individual methods perform differently across these cells that makes the combination approach a powerful technique to maximally extract information from the available data.

We next provide a comparison between the \pz PDFs computed by  the three individual techniques and the ${\rm BMC}$ technique and the SDSS spectroscopic redshift for all 1,097,397 galaxies in Figure~\ref{fig:combined_true_sdss}. The first observation from the figure is the bi-modality of the sample, which is the result of the two primary sub-populations (\ie MGS and LRGs). Overall, the results are quite good with a very tight correlation, especially in areas of high source density areas. The main exception is at the highest redshifts where there is a slight underestimation; and, as seen before, we can observe how these different approaches provide similar results, which are therefore correlated, while still differing in other areas where one method may outperform the others. The most right panel is the  ${\rm BMC}$  which shows a slightly tighter distribution in comparison to the others.

\begin{figure}
\includegraphics[width=0.48\textwidth]{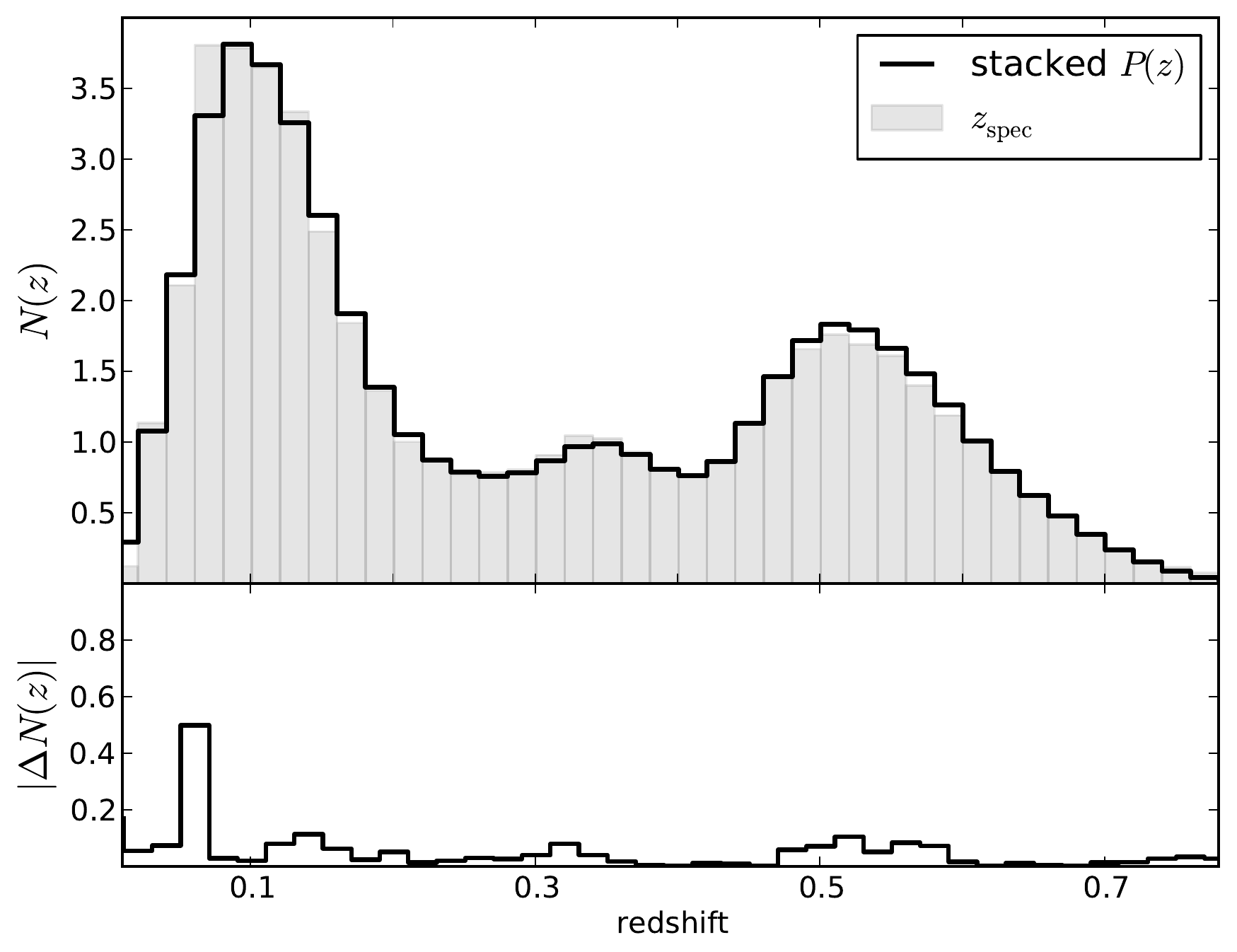}
\caption{Top panel: The $N(z)$ computed directly from the spectroscopic redshifts (gray) and by stacking the \pz PDF estimates from the $BMC$ method (black). Bottom Panel: The absolute difference between these two $N(z)$ distributions.} 
\label{fig:N_z_sdss}
\end{figure}

Finally, in Figure~\ref{fig:N_z_sdss} we present the galaxy redshift distribution for both the spectroscopic sample (in gray) and the photometric redshift distribution, computed by stacking the individual galaxy PDFs (in black). This Figure highlights that the underestimation of the \pz at high redshifts in Figure~\ref{fig:combined_true_sdss}  coincides with the strong decline in the number of galaxies after $z = 0.75$. More importantly, however, this $N(z)$ figure shows the excellent agreement between the photometric and spectroscopic galaxy redshift distributions. Given the fact that the SDSS galaxy sample contains two distinct populations, this agreement is remarkable.

\section{Outliers identification}

As we have discussed previously, aggregating information from multiple \pz PDFs estimation techniques can improve the overall \pz solution. In this section, however, we explore how this information can be combined to improve the identification of outliers within the test data. In particular, we attempt to use all possible information in order to identify these objects, from the shape of each \pz PDF as computed by all individual methods to the differences in their predicted \pzns. We adopt a Na\"{\i}ve Bayes Classifier (NBC) \citep{Zhang2004} to identify these two groups, a technique that has found widespread adoption to identify spam email messages. The advantage of this approach is that it is easy to implement, is fast and efficient for large dimensional data, and can be very competitive with other classifiers \citep{Domingos1997,Frank2000}. 

Let $\boldsymbol{\theta}$ be the set of $N_{\theta}$ parameters, $\theta_i$, we will use to identify the outliers. By using the Bayes Theorem, we can compute the probability for an object to be an outlier, given $\boldsymbol{\theta}$ as:
\begin{equation}\label{NBC1}
 P({\rm out} \mid \boldsymbol{\theta} ) = \frac{P({\rm out}) P(\boldsymbol{\theta}  \mid {\rm out})}{P(\boldsymbol{\theta})}
\end{equation}
where the \textit{evidence}, $P(\boldsymbol{\theta})$ is given by
\begin{equation}
 P(\boldsymbol{\theta}) = P(\boldsymbol{\theta}  \mid {\rm out}) + P(\boldsymbol{\theta}  \mid {\rm in})
\end{equation}
and \textit{out} refers to outliers and \textit{in} refers to inliers, the only two classes we identify in this analysis. The Na\"{\i}ve Bayes Classifier assumes that all  $\theta_i$ variables are independent, even if their independence is weak or even if there is a strong dependence between any of them. Each variable provides information about these two classes, and this information can be combined to make a stronger classifier \citep{Zhang2004}. For instance, in CB13 we showed that outliers tend to have a broader (larger values of $zConf$) and multi-peaked PDFs, and herein we treat these values as independent data even though multi-peaked PDFs are indeed generally broader. 

By using this assumption, we can write:
\begin{equation}
  P(\boldsymbol{\theta}  \mid {\rm out}) = P(\theta_1, \theta_2, \dots, \theta_{N_\theta} \mid {\rm out}) = \prod\limits_{i=1}^{N_\theta} P(\theta_i \mid {\rm out})
\end{equation}
and similarly,
\begin{equation}
  P(\boldsymbol{\theta}  \mid {\rm in}) = \prod\limits_{i=1}^{N_\theta} P(\theta_i \mid {\rm in})
\end{equation}
We can now rewrite Equation~\ref{NBC1}:
\begin{equation}\label{NBC2}
 P({\rm out} \mid \boldsymbol{\theta} ) = \frac{ P({\rm out})  \prod P(\theta_i \mid {\rm out})}{\prod P(\theta_i \mid {\rm out}) + \prod P(\theta_i \mid {\rm in}) } ,
\end{equation}
which is similar to the method used by \cite{Gorecki2014}, who demonstrated the potential of this approach to identify \pz outliers. Here, however, we use a different set of variables that are generated for all three individual \pz PDF methods. 

In our case we use $N_{peak}$, the number of peaks in each \pz PDF; $r_{peak}$, the logarithm of the ratio between the height of the first peak and the height of the second peak; $z_{mean}$, the mean of each \pz PDF; $z_{mode}$, the mode of each PDF;$zConf$, measured with respect to the mean and the mode of the \pz PDF; and the difference in the \pz, as enumerated by the mean and the mode between each of the three methods. Thus, we have six metrics computed individually for each of our three \pz PDF estimation techniques, and an additional six metrics for the difference in \pz mean and mode between the three techniques. As a result, we have a total of twenty-four metrics, to which we can add the input data for each survey. 

We, therefore, have a total of thirty-eight variables for the DEEP2 survey, while for the SDSS we have a total of thirty-three variables to use for outlier detection. For convenience, we rescale each of these variables to lie between zero and one.  $P(\theta_i \mid {\rm in})$  and $P(\theta_i \mid {\rm out})$ are evaluated by using the OOB or cross-validation data, which we have shown can reliably predict the results on the test data. Once  computed, these distributions are evaluated for the test data, where $ P({\rm out} \mid \boldsymbol{\theta} ) $ is evaluated separately for each galaxy in the test data. 

\begin{figure}
\includegraphics[width=0.48\textwidth]{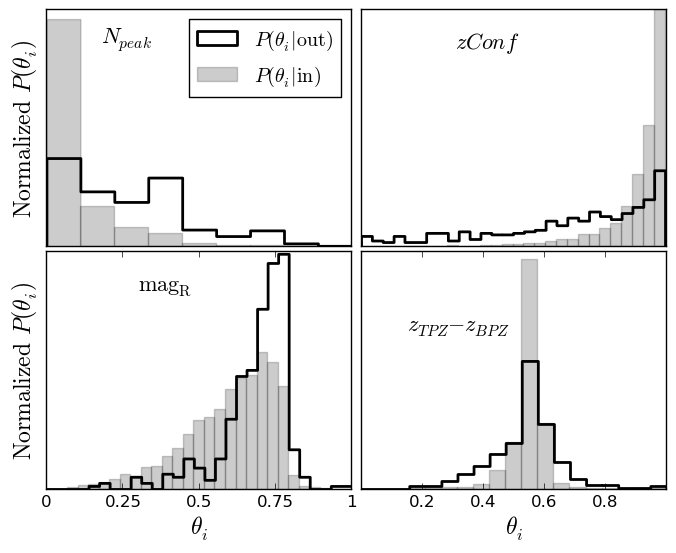}
\caption{The normalized distributions of four of the set of thirty-eight (rescaled) $\boldsymbol{\theta} $ variables from the DEEP2 data that are used for outlier detection. The variables are binned as outliers (black line histograms) or inliers (gray histogram). From the top left and following in a clockwise direction: $N_{peak}$, the number of peaks in the \tpz PDF; $zConf$, as computed from \tpzns, the $R$-band magnitude, and the difference between the \pz computed by using the mean of the \tpz and \bpz PDFs. 
\label{fig:P_theta}}
\end{figure}

\begin{figure}
\includegraphics[width=0.48\textwidth]{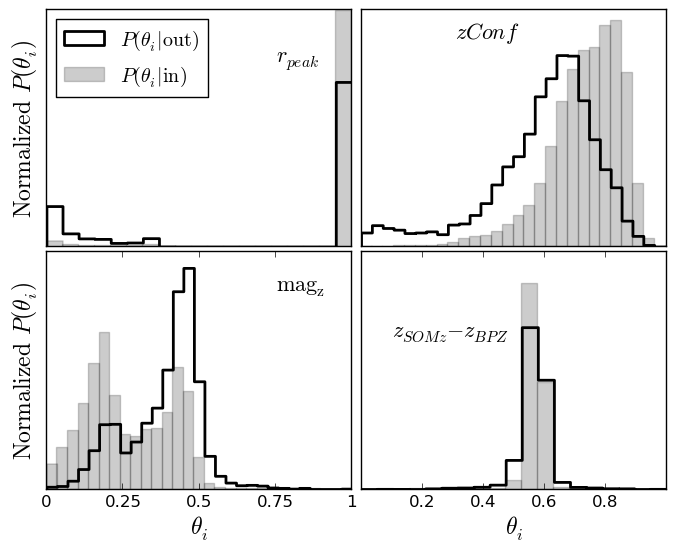}
\caption{The normalized distributions of four of the set of thirty-three (rescaled) $\boldsymbol{\theta} $ variables from the SDSS data that are used for outlier detection. The variables are binned as outliers (black line histograms) or inliers (gray histogram). From the top left and following in a clockwise direction: $r_{peak}$, the logarithmic ratio of the first two peaks in the \tpz PDF; $zConf$, as computed from \somzns, the SDSS $z$-band magnitude, and the difference between the \pz computed by using the mode of the \somz and \bpz PDFs.
\label{fig:P_theta_sdss}}
\end{figure}

Figure~\ref{fig:P_theta} presents the normalized distributions of four rescaled variables (\ie $\theta_i$) taken from the DEEP2 test data. Note that the inlier and outlier distributions are normalized to have unit area, thus these distributions illustrate how these two populations differ and not how the relative numbers between the inlier and outlier populations vary. The four variables shown in this Figure include the number of peaks in the \tpz PDFs, $zConf$ computed by \tpzns, the $R$-band magnitude, and the difference between the mean of the \tpz and \bpz \pz PDFs. In just these four distributions, there is clear separation between the galaxies labeled as outliers (black line) and inliers (gray shaded area), where the outlier identification metrics are defined by using Table~\ref{tab:def_metrics}. In particular, for this Figure we use ${\rm out}_{0.1}$, \ie galaxies for which $\Delta z' > 0.1$. While not shown, a similar result is seen for the other distributions. The result that outliers and inliers 
follow distinct distributions is what makes this a powerful approach. In effect, all information is assumed to be independent, and when combined allows an efficient identification of catastrophic outliers. 

We see a similar trend in Figure~\ref{fig:P_theta_sdss}, but now for galaxies in the SDSS test data. In this Figure, we have selected four different rescaled variables; namely, the logarithmic ratio between the first and the second peaks of the \tpz PDF (note that if the PDF has one peak, we fix this value to be four), the $zConf$ computed from \somzns, the SDSS $z$-band magnitude, and the difference between the mode of the \somz and \bpz \pz PDFs. Once again, this Figure highlights that in each of these distributions there is a separation between the outliers and inliers, and that in combination we obtain an even better discriminant between these two classes.

\begin{figure}
\includegraphics[width=0.48\textwidth]{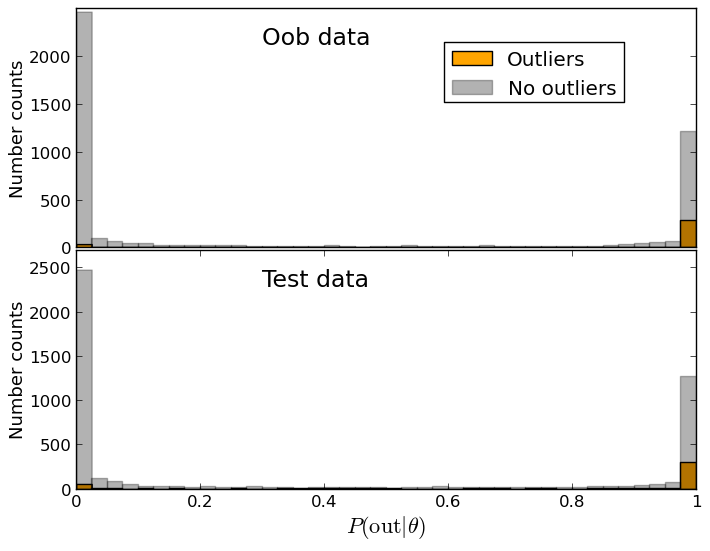}
\caption{The count distribution of $P({\rm out} \mid \boldsymbol{\theta} )$ for the DEEP2 OOB data (top) and test data (bottom) showing both the outliers (orange) and inliers (gray).
\label{fig:P_outlier_deep}}
\end{figure}

\begin{figure}
\includegraphics[width=0.48\textwidth]{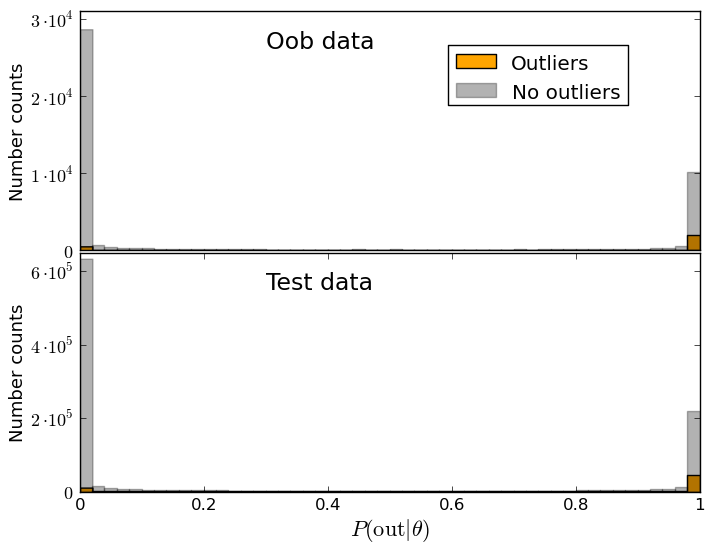}
\caption{The count  distribution of $P({\rm out} \mid \boldsymbol{\theta} )$ for the SDSS OOB data (top) and test data (bottom) showing both the outliers (orange) and inliers (gray ). 
\label{fig:P_outlier_sdss}}
\end{figure}

By using Equation~\ref{NBC2}, we can combine the values of all of the rescaled variables (\ie $\theta_i$) to compute $P({\rm out} \mid \boldsymbol{\theta} )$ for each galaxy in the DEEP2 and SDSS, both for the OOB and the test data. We present these $P({\rm out} \mid \boldsymbol{\theta} )$ distributions for the DEEP2 in Figure~\ref{fig:P_outlier_deep} and for the SDSS in Figure~\ref{fig:P_outlier_sdss}. Both Figures are similar, showing a clear separation between the outliers and inliers in both data sets. The probability ranges between zero and one, and the outliers are generally concentrated near one, while the inliers are concentrated near zero. While some mis-classifications remain, the contamination has been greatly reduced, meaning we can successfully identify a majority of the outlier population. Lastly, while there are a few galaxies with probabilities lying somewhere between zero and one, these distributions are highly bimodal, which reinforces the belief that this method provides a remarkably good 
discriminant between these two populations. 

Once again, in both Figures~\ref{fig:P_outlier_deep} and \ref{fig:P_outlier_sdss}, the OOB and test data distributions show strong similarities. As a result, we can expect that any cut we make on the OOB data will produce similar results in the test data, allowing us to make a robust classification of outliers in potentially blind test data. To quantify this, we show in Table~\ref{tab:outliers_deep} the effects of selecting outliers by using this NBC approach and by using the $zConf$ approach we initially presented in CB13 for the DEEP2 data. To simplify the comparison, we first select inlier galaxies by using the $P({\rm out} \mid \boldsymbol{\theta} )$ to cut the test data sample, and subsequently choosing those galaxies in the test data that have the highest $zConf$ so that we have the same number of galaxies selected via both techniques. 

\begin{table}
\caption{The effect of removing outliers from the DEEP2 test data on several, select performance metrics by using the Na\"{\i}ve Bayes Classifier and the $zConf$ cut approach. The two techniques are applied to ensure equal numbers of galaxies are selected, which is indicated by the \textit{Fraction} column.
\label{tab:outliers_deep}}
\centering
\renewcommand{\footnoterule}{}
\begin{tabular}{lllrrr}	
Method &Criteria &Fraction & $<\Delta z'>$ & $\sigma_{\Delta z'}$ & ${\rm out}_{0.1}$ \\
\hline
NBC & $<$ 0.998 & 83.0 \% & 0.02819 & 0.03948 & 0.0362\\
$zConf$ & $>$ 0.854 &83.0 \% & 0.02868 & 0.04186 & 0.0371\\
\hline
NBC & $<$ 0.894 &72.0 \%& 0.02616 & 0.03548 & 0.0304\\
$zConf$ & $>$ 0.893& 72.0 \%& 0.02721 & 0.03895 & 0.0330\\
\hline
NBC & $<$ 0.174& 56.0 \%& 0.02565 & 0.03470 & 0.0251\\
$zConf$ &$>$ 0.918&56.0 \%& 0.02595 & 0.03575 & 0.0289\\
\end{tabular}
\end{table}

The information in this Table demonstrates that the NBC approach produces a sample of galaxies that have a smaller spread in $\Delta z'$ along with a smaller number of outliers than the $zConf$ method, which was previously shown to be beneficial in this regard (CB13). We interpret this result as suggesting that a $zConf$ cut can potentially remove \textit{good} galaxies whose \pz PDF happens top be broad, while retaining some \textit{bad} galaxies that have a well-localized \pz PDF. By using a Na\"{\i}ve Bayes approach, we collect all information from \pz PDFs predicted by using different, semi-independent methods, allowing a more robust discriminant between outliers and inliers. Finally, we notice that as always there is a trade-off between completeness, whereby we try to retain as many \textit{good} galaxies, and contamination, whereby we try to minimize the inclusion of \textit{bad} galaxies. The final choice in this conflict should be determined by the scientific application, but by producing a 
probabilistic value, subsequent researchers can make these cuts more easily.

We performed a similar analysis on the SDSS galaxy sample and present the results in Table~\ref{tab:outliers_sdss}. As was the case with the DEEP2 galaxies, we see that the NBC approach once again does better in identifying outliers within the sample, as the NBC cuts have a smaller scatter and the fraction of remaining outliers is remarkably small. We also notice that the mean bias is similar between the two approaches, but the number of outliers, defined as $\Delta z' > 0.1$, is significantly reduced when we adopt the  Bayesian approach. This is yet another piece of evidence supporting the benefits of aggregating information to make decisions. 

\begin{table}
\caption{The effect of removing outliers from the SDSS test data on several, select performance metrics by using the Na\"{\i}ve Bayes Classifier and the $zConf$ cut approach. The two techniques are applied to ensure equal numbers of galaxies are selected, which is indicated by the \textit{Fraction} column.
\label{tab:outliers_sdss}}
\centering
\renewcommand{\footnoterule}{}
\begin{tabular}{lllrrr}
Method &Criteria &Fraction & $<\Delta z'>$ & $\sigma_{\Delta z'}$ & ${\rm out}_{0.1}$ \\
\hline
NBC & $<$ 0.999 & 83.0 \% & 0.01560 & 0.01533 & 0.0022\\
$zConf$ & $>$ 0.7018 &83.0 \% & 0.01589 & 0.01704 & 0.0035\\
\hline
NBC & $<$ 0.802 &72.0 \%& 0.01473 & 0.01411 & 0.0012\\
$zConf$ & $>$ 0.755& 72.0 \%& 0.01475 & 0.01549 & 0.0026\\
\hline
NBC & $<$ 0.001& 56.0 \%& 0.01387 & 0.01309& 0.0006\\
$zConf$ &$>$ 0.807 &56.0 \%& 0.01366 & 0.01410 & 0.0020\\
\end{tabular}
\end{table}

We can also test how the definition of an outlier affects this approach. Previously we identified an outlier as a galaxy that had $\Delta z' > 0.1$; but for the purpose of this test, we apply a much more restrictive cut of $\Delta z' > 0.05$. We apply the NBC cut and produce a matched sample by imposing a $zConf$ cut to both the DEEP2 and the SDSS galaxies, presenting the information in Table~\ref{tab:outliers_both}. We find, once again, that even for this more restrictive approach we produce a cleaner catalog (of the same size) as compared to using only the $zConf$ parameter. Interestingly, even after removing almost 30\% of the galaxies from the DEEP2 galaxy sample, we still have over a 10\% outlier contamination. On the other hand, this tight cut applied to the SDSS galaxies produces a very small contamination of $\sim$ 2\%, for both methods, albeit the NBC approach is still slightly better.

\begin{table}
\caption{The effect of removing outliers, defined as $\Delta z' > 0.05$, from the DEEP2 and SDSS test data on several, select performance metrics by using the Na\"{\i}ve Bayes Classifier and the $zConf$ cut approach. For each data set, the two techniques are applied to ensure equal numbers of galaxies are selected, which is indicated by the \textit{Fraction} column.
\label{tab:outliers_both}}
\centering
\renewcommand{\footnoterule}{}
\begin{tabular}{lllrrr}
Method &Criteria &Fraction & $<\Delta z'>$ & $\sigma_{\Delta z'}$ & ${\rm out}_{0.05}$ \\
\hline
DEEP2 &&&&&\\
\hline
NBC & $<$ 0.996 &72.0 \%& 0.02780 & 0.03934 & 0.138\\
$zConf$ & $>$ 0.878& 72.0 \%& 0.02809 & 0.04244 & 0.141\\
\hline
SDSS &&&&&\\
\hline
NBC & $<$ 0.85& 72.0 \%& 0.01461 & 0.01407& 0.0247\\
$zConf$ &$>$ 0.75 &72.0 \%& 0.01479 & 0.01554 & 0.0278\\
\end{tabular}
\end{table}

While producing galaxy samples that are less affected by outliers than competing techniques, the NBC approach has an additional advantage in that it can easily be extended to other variables and to other \pz algorithms. In effect, any information that might increase the efficacy of outlier identification can be included in order to improve this discriminant while still maximizing the overall galaxy sample size. 

\section{Conclusions}

We have presented and analyzed different techniques for combining \pz PDF estimations on galaxy samples from the DEEP2 and SDSS projects. In particular, we use three independent \pz PDF estimation methods: \tpzns, a supervised machine learning technique based on prediction trees and a random forest; \somzns, an unsupervised machine learning approach based on self organizing maps and a random atlas; and \bpzns, a standard template-fitting method that we have slightly modified to parallelize the implementation. Both \tpz and \somz are currently available within a new software package entitled \texttt{MLZ}\footnote{http://lcdm.astro.illinois.edu/code/mlz.html}.

We developed seven different combination methods that employ ensemble learning with cross-validation data to maximize the information extracted. Of these seven methods, four employ a weighted average where the weights can either be selected to be uniform across the input methods, to be determined from the shape of the \pz PDF (e.g., by using the $zConf$ parameter), to be determined by an \textit{oracle} estimator where one  (ideally the best) method is preferentially selected, and where the weights are obtained by a fitting procedure applied to the OOB data. Three of the combination methods were Bayesian techniques: Bayesian Model Averaging (BMA), Bayesian Model Combination (BMC), and Hierarchical Bayes (HB).

We expect the individual \pz PDF estimation techniques to perform differently across the parameter space spanned by our galaxy samples; for example, template-fitting techniques are expected to work better at higher redshifts than machine learning methods, which perform optimally when provided high-quality, representative training data. Thus we construct a two-dimensional, $10 \times 10$ self-organizing map (SOM) to subdivide the high-dimensional parameter space occupied by the galaxy samples. We apply different \pz PDF estimation techniques within each cell in this map, since each cell should contain galaxies with similar properties. A visual inspection of these maps indicates that the two machine learning methods: \tpz and \somz are generally complementary, and that in combination with a model based technique such as \bpz we are able to maximize the coverage of this multidimensional space  efficiently.

We also verified that by using the OOB data, as introduced in CB13, we can an obtain an accurate, unbiased and \textit{honest} estimation of the performance of a \pz PDF estimation technique on the test data. We also computed the correlation coefficient and the error distribution and showed they also behave similarly for the cross-validation (\ie the OOB data) and the test data. These computations are extremely important when combining \pz PDF techniques as we can learn from the OOB data the optimal parameters needed for a specific ensemble learning approach, and thereby maximize the performance of that combination technique when applied to \textit{blind} test data.
 
Overall, we found that the BMA and BMC are the best  \pz PDF combination techniques as they have better performance metrics when compared to the individual \pz PDF estimation techniques, especially when unbiased cross-validation data is available. This result is true for both the DEEP2 and the SDSS data. When OOB data is not available, we can instead use the $zConf$ parameter as a weight for each method after first renormalizing the individual \pz PDFs. We can also use the Hierarchical Bayes method to combine these predictions, which we demonstrated can also lead to better results.

Within this Bayesian Framework, we also developed a novel, Na\"{\i}ve Bayesian Classifier (NBC) that efficiently identifies outliers within the galaxy sample. The approach we present gathers all available information from the different \pz PDF estimation techniques regarding the shape of the PDF, the location of the mean and mode, and the magnitudes and colors, which are all  \textit{naively} assumed to be independent, in order to compute a Bayesian posterior probability that a certain galaxy is an outlier. The distribution of these probabilities for an entire galaxy sample indicate that this is a very powerful method to separate outliers from inliers (\ie \textit{good} galaxies), and we further demonstrated that this approach can produce a more accurate and cleaner sample of galaxies than competing techniques, such as the use of the $zConf$ parameter. An important takeaway point is that all information provided by the catalogs and the \pz PDF methods, no matter how redundant the information might appear, 
helps in building this discriminant probability. Given the probabilistic nature of this computation, the final application of this technique can be chosen to maximize the scientific utility of the resulting galaxy data for a specific application. 

The computational cost to apply these Bayesian models to galaxy samples will depend directly on the size of the data set, the number of \pz estimation techniques used, and the resolution of the given \pz PDFs. In~\cite{CarrascoKind2014b} we demonstrate how a sparse basis representation can reduce the storage significantly and that manipulation of these PDFs can be improved within the bases framework thereby reducing computational costs. We plan to adopt this representation framework to compute the combination models, which will allow fast and accurate combination of multiple \pz PDFs.

Finally, we have demonstrated that even when a \pz PDF technique is very accurate, we can still make improvements by extracting additional information about the distribution of  galaxies in the higher dimensional parameter space and the individual performance of the \pz PDF algorithms. There are currently a large number of published algorithms to compute \pz's, many of which also compute \pz PDFs. Even if their performance is similar, these techniques will all have their own advantages and disadvantages. Thus we believe the combination of different techniques is the future of \pz research, and we expect additional research to be forthcoming in this area. Overall, the combination of \pz PDFs is a powerful, new approach that can be easily extended to incorporate new techniques in order to generate a meta-predictor that accelerate our knowledge and understanding of the Universe.

\section*{Acknowledgements}
The authors thank the referee for a careful reading of the manuscript and for comments that improved this work. RJB and MCK acknowledge support from the National Science Foundation Grant No. AST-1313415. MCK has been supported by the Computational Science and Engineering (CSE) fellowship at the University of Illinois at Urbana-Champaign. RJB has been supported in part by the Institute for Advanced Computing Applications and Technologies faculty fellowship at the University of Illinois.

The authors gratefully acknowledge the use of the parallel computing resource provided by the Computational Science and Engineering Program at the University of Illinois.  The CSE computing resource, provided as part of the Taub cluster, is devoted to high performance computing in engineering and science. This work also used resources from the Extreme Science and Engineering Discovery Environment (XSEDE), which is supported by National Science Foundation grant number OCI-1053575.

Funding for the DEEP2 Galaxy Redshift Survey has been provided by NSF grants AST-95-09298, AST-0071048, AST-0507428, and AST-0507483 as well as NASA LTSA 
grant NNG04GC89G.

Funding for SDSS-III has been provided by the Alfred P. Sloan Foundation, the Participating Institutions, the National Science Foundation, and the U.S. Department of Energy Office of Science. The SDSS-III web site is http://www.sdss3.org/.

SDSS-III is managed by the Astrophysical Research Consortium for the Participating Institutions of the SDSS-III Collaboration including the University of Arizona, the Brazilian Participation Group, Brookhaven National Laboratory, Carnegie Mellon University, University of Florida, the French Participation Group, the German Participation Group, Harvard University, the Instituto de Astrofisica de Canarias, the Michigan State/Notre Dame/JINA Participation Group, Johns Hopkins University, Lawrence Berkeley National Laboratory, Max Planck Institute for Astrophysics, Max Planck Institute for Extraterrestrial Physics, New Mexico State University, New York University, Ohio State University, Pennsylvania State University, University of Portsmouth, Princeton University, the Spanish Participation Group, University of Tokyo, University of Utah, Vanderbilt University, University of Virginia, University of Washington, and Yale University.
\bibliographystyle{mn2e}
\bibliography{combine_bayes_final}

\bsp
\label{lastpage}
\end{document}